\documentclass[a4paper,11pt]{article}

\usepackage{jheppub} 

\usepackage{graphicx}

\usepackage{amsmath}
\usepackage[utf8]{inputenc}

\usepackage{amssymb}
\usepackage{amsfonts}
\usepackage{amsbsy}
\usepackage{url}
\usepackage{enumerate}
\usepackage{caption}
\usepackage{color}
\usepackage{ulem}
\usepackage{bm}
\usepackage{multirow,bigdelim}

\newcommand{\be}{\begin{eqnarray}}
\newcommand{\ee}{\end{eqnarray}}
\newcommand{\p}{\partial}

\makeatletter
\@addtoreset{equation}{section}
\makeatother

\newcommand\rsout{\bgroup\markoverwith{\textcolor{red}{\rule[0.5ex]{2pt}{0.4pt}}}\ULon}

\title{
Exact solutions of domain wall junctions in arbitrary dimensions
}

\author[a,b]{Minoru Eto,}
\author[a]{Masaki Kawaguchi,}
\author[b,c]{Muneto Nitta,}
\author[a]{Ryotaro Sasaki}

\affiliation[a]{Department of Physics, Yamagata University,\\
Kojirakawa-machi 1-4-12, Yamagata, Yamagata 990-8560, Japan}
\affiliation[b]{Research and Education Center for Natural
Sciences, Keio University, Hiyoshi 4-1-1, Yokohama, Kanagawa 223-8521, Japan}
\affiliation[c]{
Department of Physics, Keio University, Hiyoshi 4-1-1, Yokohama, Kanagawa 223-8521, Japan}

\emailAdd{meto(at)sci.kj.yamagata-u.ac.jp}
\emailAdd{ddwbb.daigaku(at)gmail.com}
\emailAdd{nitta(at)phys-h.keio.ac.jp}

\abstract{
Exact analytic solutions of static, stable, non-planar BPS domain wall junctions 
are obtained in extended Abelian-Higgs models in $(D+1)$-dimensional spacetime. 
For specific choice of mass parameters,
the Lagrangian is invariant under the symmetric group ${\cal S}_{D+1}$ of degree $D+1$  
spontaneously broken down to ${\cal S}_D$ in vacua,
admitting ${\cal S}_{D+1}/{\cal S}_D$ domain wall junctions.
In $D=2$, there are three vacua and three domain walls meeting at a junction point, 
in which the conventional topological charges $Y$ and $Z$ exist for 
the BPS domain wall junctions and the BPS domain walls, respectively
as known before.
In $D=3$, there are four vacua, 
six domain walls, four junction lines on which three domain walls meet, 
and one junction point on which all the six domain walls meet.  
We define a new topological charge $X$ for the junction point 
in addition to the conventional topological charges $Y$ and $Z$. 
In general dimensions, 
we find that the configuration expressed in the $D$-dimensional real space is dual to
a regular $D$-simplex in the $D$-dimensional internal space and that
a $d$-dimensional subsimplex of the regular $D$-simplex corresponds to a $(D-d)$-dimensional intersection. 
Topological charges are generalized to the level-$d$ wall charge $W_d$ for the $d$-dimensional subsimplexes.
}
\preprint{YGHP-20-01}

\begin{document}
\maketitle


\section{Introduction}
\label{sec:intro}

Domain walls (or kinks) are the simplest topological solitons separating 
discrete vacua or ground states 
\cite{Manton:2004tk,Vachaspati:2006zz,Rajaraman:1982is}, 
often created in phase transitions associated with 
spontaneous breakings of discrete symmetries~\cite{Kibble:1976sj,Kibble:1980mv}
in various systems from small to large 
such as 
magnets \cite{magnetism}, 
graphenes \cite{graphene}, 
carbon nanotubes, 
chiral p-wave superconductors \cite{chiral-p-wave}, 
Bose-Einstein condensations of ultracold atomic gases 
\cite{Takeuchi:2012ee},
helium superfluids~\cite{volovik,Zurek:1985qw,Zurek:1996sj}, 
nuclear matter 
\cite{Eto:2012qd,Yasui:2019vci} 
as well as quark matter 
\cite{Eto:2013hoa} relevant for interior of neutron stars, 
and our Universe~\cite{Vilenkin:2000jqa,Kibble:1980mv}. 
In cosmology, if they appear at a phase transition in the early Universe, 
then there happens the so-called domain wall problem \cite{Vilenkin:2000jqa}, 
that is, the domain wall energy dominates Universe to make it collapse. 
However, if the tension of the domain walls is sufficiently low, 
cosmological domain wall networks are allowed
and are suggested as a candidate of dark matter and/or dark energy 
\cite{cosmology}.
In helium superfluids, such domain walls are created in a similar manner, 
thereby simulating cosmological phase transitions~\cite{Zurek:1985qw,Zurek:1996sj}.

On the other hand,
it is widely known that supersymmetry (SUSY) is very intimate notion with
various topological solitons such as domain walls, vortices, monopoles and instantons etc 
\cite{Shifman:2009zz}.
As one of fascinating features, SUSY allows for topological solitons to 
be so-called Bogomol'nyi-Prasad-Sommerfield (BPS) states 
\cite{Bogomolny:1975de,Prasad:1975kr} 
which attain the minimum energy for a fixed boundary condition
or topology.
The BPS solitons satisfy first order differential equations, 
so-called BPS equations,
rather than equations of motion which are of second order differential equations, 
and they preserve a fraction of SUSY.
Their topological charges are directly connected to central charges of 
SUSY algebras.
The BPS domain walls in $3+1$ dimensional spacetime have been extensively
studied in ${\cal N}=1$ SUSY theories \cite{Dvali:1996xe,Dvali:1996bg,Kovner:1997ca,Smilga:1997pf,Chibisov:1997rc,Kaplunovsky:1998vt,deCarlos:1999xk,Dvali:1999pk,Edelstein:1997ej,Naganuma:2000gu,Arai:2009jd,Arai:2011gg,Nitta:2014pwa,Gudnason:2016frn,Lee:2017kaj, Arai:2018tkf} 
and ${\cal N}=2$ SUSY theories 
\cite{Abraham:1992vb,Abraham:1992qv,Gauntlett:2000bd,Gauntlett:2000ib,Tong:2002hi,Lee:2002gv,Arai:2002xa,Arai:2003es,Losev:2003gs,Isozumi:2003rp,Shifman:2003uh,Isozumi:2004jc,Isozumi:2004va,Eto:2004vy,Eto:2005wf, Eto:2005cc,Eto:2006mz,Eto:2008dm}  
(see Refs.~\cite{Tong:2005un,Eto:2006pg,Shifman:2007ce,Shifman:2009zz} as a review).
They preserve a half of SUSY and therby are called $\frac{1}{2}$ BPS states,
and the corresponding SUSY central (tensorial) charge 
\cite{deAzcarraga:1989mza,Kovner:1997ca,Chibisov:1997rc} 
is conventionally expressed as $Z$.
In the models with three or more discrete vacua, there can appear multiple domain walls.
The multiple domain walls remain as the $\frac{1}{2}$ BPS states as long as they are all parallel. 

In general, it is more natural that the multiple domain walls are not parallel.
If all the domain walls have one spatial dimension (say the $z$-axis) in common, 
domain walls extend to two dimensional space (the $x$-$y$ plane). 
The domain walls meet at a line to form a domain wall junction. 
We call these two-dimensional (of codimensions two) configurations 
planar domain wall junctions.
In SUSY models, the planar domain wall junctions preserve a quarter of supersymmetry \cite{Gibbons:1999np,Carroll:1999wr,Gorsky:1999hk}. 
They are called the $\frac{1}{4}$ BPS states and
are accompanied with a junction topological charge $Y$ 
in addition to $Z_m$ ($m =1,2$).\footnote{
In ${\cal N}=2$ SUSY theories, there are other $\frac{1}{4}$ BPS states, 
vortex strings ending on domain walls 
 called  D-brane solitons \cite{Gauntlett:2000de,Shifman:2002jm,Isozumi:2004vg,Sakai:2005sp,Auzzi:2005yw,Eto:2006pg,Eto:2008mf},
instantons inside vortices \cite{Eto:2004rz},  
and intersecting vortices \cite{Naganuma:2001pu}. 
See Ref.~\cite{Eto:2005sw} for classification of all possible 
$\frac{1}{4}$ and $\frac{1}{8}$ BPS solitons.
}
The $\frac{1}{4}$ BPS domain wall junctions 
have been also studied in ${\cal N}=1$ SUSY models 
\cite{Abraham:1990nz,Gabadadze:1999pp,Oda:1999az,Shifman:1999ri,Ito:2000zf,Binosi:1999vb,Nam:2000qe,Carroll:1999mu,Naganuma:2001br,Nitta:2014pwa} 
and ${\cal N}=2$ SUSY models 
\cite{Kakimoto:2003zu,Eto:2005cp,Eto:2005fm,Eto:2005mx,Eto:2006bb,Eto:2007uc,Fujimori:2008ee,Shin:2018chr,Shin:2019buk,Kim:2019jzo}.
The domain wall junctions are similar to vortex strings but the junction charge $Y$ was found
to {\it negatively} contribute to the total energy \cite{Oda:1999az,
Shifman:1999ri,Ito:2000zf} 
in contrast to the domain wall charge $Z_{1,2}$ always contributing 
positively to the total energy.
Thus, the junction charge $Y$ should be understood as a sort of binding energy bonding the domain walls 
\cite{Oda:1999az,Ito:2000zf} rather than independent topological solitons.
Planer network of the domain walls and junctions were studied as non-BPS states
in Refs.~\cite{Saffin:1999au,Bazeia:1999xi}. Then, it was found that the ${\cal N}=2$ SUSY gauge models
can have any kind of planar domain wall networks as 
the $\frac{1}{4}$ BPS states \cite{Eto:2005cp,Eto:2005fm}, 
similarly to D-brane networks \cite{Sen:1997xi,Aharony:1997bh}.

In this paper, we study non-planar BPS domain wall junctions in which  
three or more domain walls having angles meet at a point.
Namely, we consider the domain wall junctions which are essentially $D$-dimensional in $D+1$ dimensional spacetime for $D\ge3$.
The planar domain wall junctions 
are $N$-pronged junctions ($N\ge 3$) of codimension two
typically appearing when
the models under consideration possess a $\mathbb{Z}_N$ symmetry 
spontaneously broken in the vacua, see for example 
\cite{Abraham:1990nz,Gibbons:1999np,Carroll:1999wr,Gabadadze:1999pp,Shifman:1999ri}. 
Note that the $\mathbb{Z}_N$ symmetry is an Abelian group naturally associated
with a discrete rotation group of a regular $N$-gon in two dimensions. 
The $\mathbb{Z}_N$ is a subgroup of $SO(2)$ rotation of the two dimensional space. 

In this work, we generalize this 
to the higher dimensions.
A symmetry group preserving 
a $D$-dimensional object ($D\ge3$) is usually non-Abelian since it is a
subgroup of $SO(D)$. For instance, the symmetry group of
a regular tetrahedron is the symmetric group of degree four ${\cal S}_4$ 
which is non-Abelian.
Inspired by the ${\cal N}=2$ SUSY QED in $3+1$ dimensions,
we study a $U(1)$ gauge theory coupled with $N_F$ charged scalars and $N_F'$ reals scalars 
 in $(D+1)$-dimensional spacetime. 
It turns out that the vacuum structure of the model is indeed $N_F'$ dimensional. 
Then, to obtain BPS $D$-dimensional
domain wall junctions, we set $N_F' = D$ and derive the BPS equations.
Interestingly, the BPS equations involve arbitrary $D$ signs $\xi_m=\pm 1$ ($m=1,2,\cdots,D$).
Therefore, there exist $2^D$ different sets of the BPS equations according to choices of $\xi_m$.
When $D=2$, there are four sets and we find they are identical to four sets of 
$\frac{1}{4}$ BPS equations in ${\cal N}=2$ SQED studied in
Refs.~\cite{Kakimoto:2003zu,Eto:2005cp,Eto:2005fm,Eto:2005mx,Eto:2006bb,Eto:2007uc,Fujimori:2008ee}.
In the SUSY context, the number four comes from the number of way of selecting two among eight supercharges.

The main result of this paper is to present exact analytic solutions
of non-planar domain wall junctions in generic $D+1$-dimensional spacetime. For this purpose,
we will restrict ourself to the models of $N_F=D+1$ with the largest symmetry, that is the symmetric group ${\cal S}_{D+1}$.
The previously  known exact solution of the 3-pronged planar domain wall junction \cite{Kakimoto:2003zu}
corresponds to the case of $D=2$. 
In the vacua, ${\cal S}_3$ is spontaneously broken to ${\cal S}_2$, so that
the vacuum structure is the coset ${\cal S}_3/{\cal S}_2$ 
consists of three elements.
We then construct a novel exact solution of the three-dimensional domain wall junction
connecting the four different vacua
for $D=3$.
The model has the ${\cal S}_4$ symmetry 
spontaneously broken to ${\cal S}_3$. 
The vacua are the coset
${\cal S}_4/{\cal S}_3$ with 
four elements. Reflecting the fact that 
the symmetric group ${\cal S}_4$ is the symmetry group of the regular tetrahedron, the vacua correspond to
four vertices of the tetrahedron. 
The configuration
in the real space has the same symmetric property ${\cal S}_4/{\cal S}_3$ as the vacua structure;
it consists of  six domain walls.
Two domain walls arbitrary chosen from them glue along a line junction. 
There exist four such junction lines, and all the junction lines meet at one point.
Correspondingly, the topological charges of the domain walls and domain wall junction lines are $Z_m$ and $Y_{mn}$
($m,n=1,2,3$ and $m > n$), respectively as before.
In addition, there is 
a new topological charge $X$ for the junction point. 
Although the topological charges $Z_m$ and $Y_{mn}$ contribute to the energy density,
the new topological charge $X$ 
does not. 
We then 
construct an exact solution of a BPS ${\cal S}_{D+1}/{\cal S}_D$ domain wall junction in $D+1$-dimensional spacetime.
We find several geometric properties of the solution.  
The ${\cal S}_{D+1}/{\cal S}_D$ domain wall junction expressed in the real space is dual to
a regular $D$-simplex in the internal space whose $D+1$ vertices correspond to the vacua.
A $d$-face, a $d$-dimensional subsimplex ($0 \le d \le D$) 
of the $D$-simplex (the 0-faces are the vertices, the 1-faces are the edges,
and so on) are dual to $D-d$ dimensional building blocks of the configuration.
For example, the 0-faces is dual to the $D$-dimensional vacuum domains, 
the 1-faces to the $(D-1)$-dimensional domain walls, and so on.
For each $d$, we define a topological charge $W_d$ of the level-$d$ 
($W_{1,m}=Z_m$, $W_{2,mn}=Y_{mn}$, and $W_{3,lmn}=X_{lmn}$ and so on).  
The symmetric group ${\cal S}_{D+1}$ which is the symmetry group of 
the regular $D$-simplex is isomorphic to the Coxeter group of the type $A_D$.
The orthographic projection in two dimensional plane of the $D$-simplex is known as the Coxeter plane of type $A_D$.
We find that the exact solution of the ${\cal S}_{D+1}/{\cal S}_D$ domain wall junction 
provides the Coxeter plane of the type $A_D$ when it is expressed in 
a two dimensional internal space.
 
This paper is organized as follows.
In Sec.~\ref{sec:model}, we present our model, its symmetry structure and vacua.
In Sec.~\ref{sec:BPS_eq}, we derive BPS equations for domain wall junctions. 
In Sec.~\ref{sec:A3_D2}, we construct a planar  ${\cal S}_3/{\cal S}_2$ domain wall junction in $D=2$, 
which is essentially a review of Ref.~\cite{Kakimoto:2003zu}.
In Sec.~\ref{sec:A4_D3}, we construct a ${\cal S}_4/{\cal S}_3$ domain wall junction in $D=3$. 
In Sec.~\ref{sec:simpler}, they are generalized to 
${\cal S}_4/{\cal S}_3$ domain wall junctions in $D$ dimensions, 
and geometric properties are discussed. 
Sec.~\ref{sec:conclusion} is devoted to a summary and discussion.
Appendix \label{sec:app} summarizes explicit expression of 
the symmetric group ${\cal S}_4$ and the coset ${\cal S}_4/{\cal S}_3$.


\section{The model, symmetry and vacua}
\label{sec:model}

\subsection{The model}
We study a $U(1)$ gauge theory with
$N_F$ charged complex scalar fields $H^A$ ($A = 1,2,\cdots,N_F$) and $N'_F$ real scalar fields
$\Sigma^{A'}$ ($A'=1,2,\cdots,N_F'$) in $D+1$-dimensional spacetime.
The Lagrangian is given by
\be
{\cal L} &=& -\frac{1}{4e^2}F_{\mu\nu}F^{\mu\nu} 
+ \frac{1}{2e^2}\sum_{A'=1}^{N_F'}\partial_\mu \Sigma^{A'} \partial^\mu \Sigma^{A'}
+ D_\mu H (D^\mu H)^\dagger  \\\nonumber
&\ &-~ \frac{1}{2e^2}Y^2 
- \sum_{A'=1}^{N_F'}\left(\Sigma^{A'}H - H M^{A'}\right) \left(\Sigma^{A'}H - H M^{A'}\right)^\dagger,
\label{eq:lag}
\ee
where $H$ is an $N_F$ component row vector made of $H^A$,
\be
H = \left(H^1,\ H^2,\ \cdots, H^{N_F}\right),
\ee
$Y$ is a scalar quantity defined by
\be
Y = e^2\left(v^2 - H H^\dagger\right),
\ee
and $M^{A'}$ ($A' =1,\cdots,N_F'$) are 
$N_F$ by $N_F$ real diagonal mass matrices defined by
\be
M^{A'} = {\rm diag}\left(m_{A',1},\ m_{A',2},\ \cdots,\ m_{A',N_F}\right).
\ee
The spacetime index $\mu$ runs from $0$ to $D$, and $F_{\mu\nu}$ is 
a $U(1)$ gauge field strength.
The coupling constants in the Lagrangian in Eq. (\ref{eq:lag}) are taken 
to be the so-called Bogomol'nyi limit.\footnote{This does not immediately imply that 
the model can be made supersymmetric by adding fermions.
Only in certain cases, this can be supersymmetric.}

Let us discuss the symmetry structure of our model.
When all the mass matrices are proportional to the unit matrix, the flavor symmetry for $H$ is $SU(N_F)$.
It reduces to a subgroup according to degeneracy of the mass eigenvalues.
Let $\bm{m}_A$ be an $N_F'$ vector whose components are the $A$th diagonal elements of $M^{A'}$s, namely,
\be
\bm{m}_A = \left(m_{1,A},\ m_{2,A},\ \cdots,\ m_{N_F',A}\right).
\ee
Then, the flavor symmetry $SU(N_F)$ is explicitly broken maximally 
to $U(1)^{N_{\rm F}-1}$ when
\be
\bm{m}_A \neq \bm{m}_B,\qquad \text{if } A\neq B.
\ee

When all the mass matrices are zero, there is a flavor symmetry $O(N_F')$ acting on 
$\bm{\Sigma} = \left(\Sigma^1,\ \Sigma^2,\ \cdots,\ \Sigma^{N_F'}\right)$.

When the masses are specially tuned, a discrete symmetry appears. 
This can be found by studying the following term in the potential
\be
\sum_{A'} \Sigma^{A'} H M^{A'} H^\dagger = \sum_{A',A} \Sigma^{A'} m_{A',A} |H^A|^2 
= \bm{\Sigma}^T {\cal M}\, \overrightarrow{|H|^2},
\label{eq:M}
\ee
where ${\cal M}$ is an $N_F'$ by $N_F$ matrix defined by $({\cal M})_{A',A} = m_{A',A}$,
and $\overrightarrow{|H|^2}$ is an $N_F$ vector defined by $\overrightarrow{|H|^2} = (|H^1|^2,\ |H^2|^2,\
\cdots,\ |H^{N_F}|^2)$.
Thus, a transformation $\bm{\Sigma} \to U_\Sigma \bm{\Sigma}$, $\overrightarrow{|H|^2} \to U_H \overrightarrow{|H|^2}$
satisfying
\be
U^T_{\Sigma} {\cal M} U_H = {\cal M},
\label{eq:discrete_sym}
\ee
is an extra symmetry of ${\cal L}$.

To illustrate such a discrete symmetry, 
let us give a concrete example for the case of $N_F = 3$ and $N_F' = 2$ with
\be
M^{A'=1} = m~{\rm diag}\left(1,\ \cos\frac{2\pi}{3},\ \cos\frac{4\pi}{3}\right),\quad
M^{A'=2} = m~{\rm diag}\left(0,\ \sin\frac{2\pi}{3},\ \sin\frac{4\pi}{3}\right),
\label{eq:mm_Z3}
\ee
which gives
\be
\bm{m}_1 = m \left(1,\ 0\right),\quad
\bm{m}_2 = m \left(\cos\frac{2\pi}{3},\ \sin\frac{2\pi}{3}\right),\quad
\bm{m}_3 = m \left(\cos\frac{4\pi}{3},\ \sin\frac{4\pi}{3}\right).
\label{eq:mv_Z3}
\ee
This can equally be expressed in the matrix form as
\be
{\cal M} = m\left(
\begin{array}{ccc}
1 & \cos\frac{2\pi}{3} & \cos\frac{4\pi}{3}\\
0 & \sin\frac{2\pi}{3} & \sin\frac{4\pi}{3}
\end{array}
\right).
\ee
One can easily check that ${\cal M}$ is invariant under a set of the following transformations
\be
\left[U_\Sigma,U_H\right] \in
\left\{
\begin{array}{c}
\left[
\left(
\begin{smallmatrix}
1 & 0\\
0 & 1
\end{smallmatrix}
\right),
\left(
\begin{smallmatrix}
1 & 0 & 0\\
0 & 1 & 0\\
0 & 0 & 1
\end{smallmatrix}
\right)
\right],
\left[
\left(
\begin{smallmatrix}
\cos\frac{2\pi}{3} & \sin\frac{2\pi}{3}\\
\sin\frac{2\pi}{3} & -\cos\frac{2\pi}{3}
\end{smallmatrix}
\right),
\left(
\begin{smallmatrix}
0 & 1 & 0\\
1 & 0 & 0\\
0 & 0 & 1
\end{smallmatrix}
\right)\right],\\
\left[
\left(
\begin{smallmatrix}
1 & 0\\
0 & -1
\end{smallmatrix}
\right),
\left(
\begin{smallmatrix}
1 & 0 & 0\\
0 & 0 & 1\\
0 & 1 & 0
\end{smallmatrix}
\right)\right],
\left[
\left(
\begin{smallmatrix}
\cos\frac{4\pi}{3} & \sin\frac{4\pi}{3}\\
\sin\frac{4\pi}{3} & -\cos\frac{4\pi}{3}
\end{smallmatrix}
\right),
\left(
\begin{smallmatrix}
0 & 0 & 1\\
0 & 1 & 0\\
1 & 0 & 0
\end{smallmatrix}
\right)\right],\\
\left[
\left(
\begin{smallmatrix}
\cos\frac{2\pi}{3} & -\sin\frac{2\pi}{3}\\
\sin\frac{2\pi}{3} & \cos\frac{2\pi}{3}
\end{smallmatrix}
\right),
\left(
\begin{smallmatrix}
0 & 0 & 1\\
1 & 0 & 0\\
0 & 1 & 0
\end{smallmatrix}
\right)\right],
\left[
\left(
\begin{smallmatrix}
\cos\frac{4\pi}{3} & -\sin\frac{4\pi}{3}\\
\sin\frac{4\pi}{3} & \cos\frac{4\pi}{3}
\end{smallmatrix}
\right),
\left(
\begin{smallmatrix}
0 & 1 & 0\\
0 & 0 & 1\\
1 & 0 & 0
\end{smallmatrix}
\right)\right]
\end{array}
\right\}.
\ee
It is obvious that $\{U_H\}$ is the complete set of the symmetric group of degree three ${\cal S}_3$, 
which is the group of all permutations of a three-element set. Moreover, $\{U_\Sigma\}$ is 
the 2 by 2 matrix representation of ${\cal S}_3$.
We should verify if the other terms in the Lagrangian is invariant or not.
Clearly, both $HH^\dagger$ and $D_\mu H (D^\mu H)^\dagger$ are invariant under any transformations
$\{U_H\}$.
Then, $\sum_{A'} H M_{A'}{}^2 H^\dagger$ is the only term which we need to check.
For the special mass matrix given in Eq.~(\ref{eq:mm_Z3}), we have
\be
\sum_{A'=1}^2 H M_{A'}{}^2 H^\dagger = m^2 HH^\dagger,
\label{eq:HMMH}
\ee
and so it is also invariant.
Henceforth, the global symmetry of the Lagrangian is $U(1)^2 \times {\cal S}_3$ in this special case.\footnote{
In Ref.~\cite{Kakimoto:2003zu}, the symmetry is said as $\mathbb{Z}_3$ but it is indeed ${\cal S}_3$.}
This kind of discrete symmetry will play an important role when we construct an exact solution
of a domain wall junction.

\subsection{Vacua}
\begin{figure}[h]
\begin{center}
\includegraphics[width=15cm]{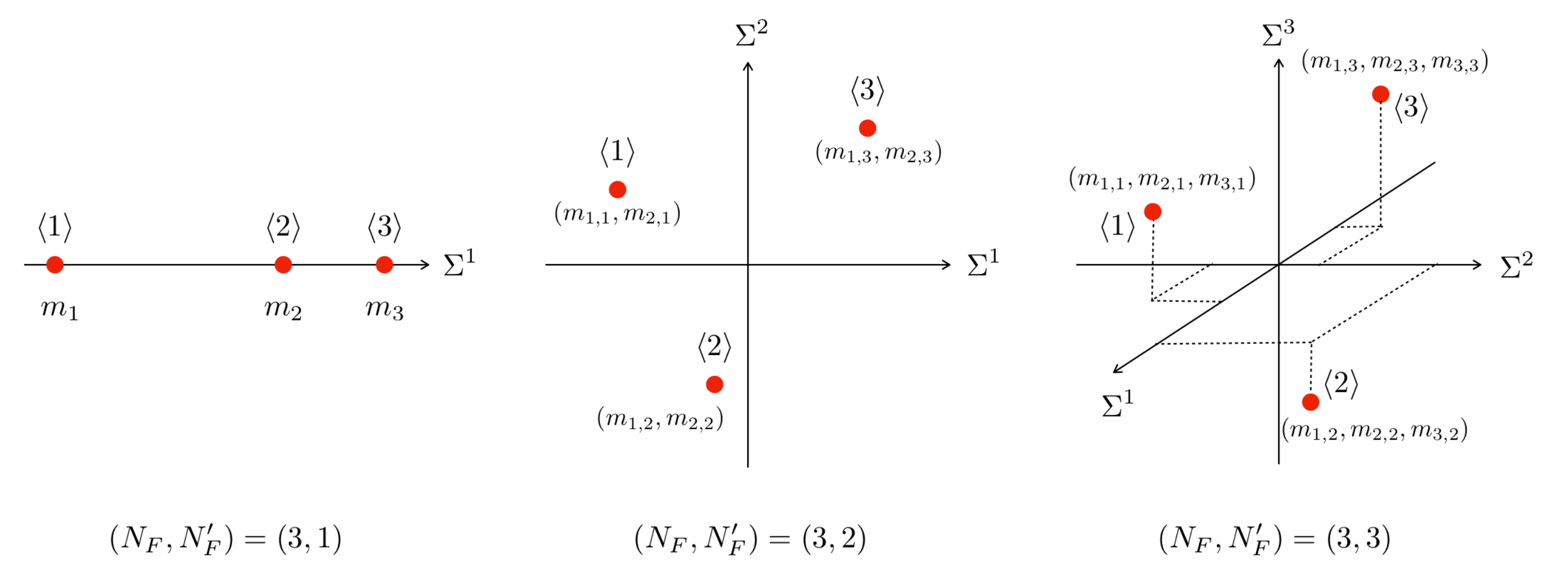}
\caption{The discrete vacua correspond to isolated points (red circles) 
in the internal $\bm{\Sigma}$ space. We show three examples
with $N_F = 3$ and $N_F'=1,2,3$.}
\label{fig:vacua}
\end{center}
\end{figure}

Since the scalar potential is positive semidefinite, a classical vacuum of the theory is determined by $V=0$.
Thus, the vacuum condition reads
\be
HH^\dagger = v^2,\qquad \Sigma^{A'} H - H M^{A'} = 0.
\label{eq:vac_cond}
\ee
In general, there are $N_F$ discrete vacua given by
\be
\left<A\right> :\quad H^B = v \delta^B_A,\qquad \Sigma^{B'} = m_{B',A}.
\label{eq:vac}
\ee
While we can specify the vacua by using either $H^A$ or $\Sigma^{A'}$, 
it will turn out that
$\Sigma^{A'}$ is more useful and so we express the vacua in the $N_F'$ dimensional internal space
spanned by $\Sigma^{A'}$.
The vacua are identical to the discrete points in the $\Sigma$ space. The
$\left<A\right>$ vacuum corresponds to the point specified by
\be
\left<A\right> :\bm{\Sigma} = \bm{m}_A,
\ee
where $\bm{m}_A$ is the $N_F'$ vector whose components are the $A$th eigenvalues of $M^{A'}$, namely,
$\bm{m}_A = \left(m_{1,A},\ m_{2,A},\ \cdots,\ m_{N_F',A}\right)$.
Hence, the number of the discrete vacua depends only on $N_F$. Fig.~\ref{fig:vacua} shows three examples
with $N_F=3$ and $N_F' = 1,2,3$.
Note that the vacua have an $N_F'$ dimensional structure when $N_F \ge N_F'$.

\subsection{Comments on supersymmetry}
Note that the parameters of the above Lagrangian are tuned in such a way that
it becomes identical to a bosonic part of supersymmetric Lagrangian.
For comparison, let us write down the ${\cal N}=2$ supersymmetric Lagrangian in $3+1$ dimensions:
\be
{\cal L}^{(D=3)}_{{\cal N}=2} &=& 
- \frac{1}{4e^2}F_{\mu\nu}F^{\mu\nu} + \frac{1}{2e^2}\sum_{A'=1}^2\partial_\mu \Sigma^{A'}\partial^\mu \Sigma^{A'} 
+ \sum_{i=1}^2D_\mu H_i (D^\mu H_i )^\dagger
\\\nonumber
&\ &-~ \frac{1}{2e^2}\sum_{a=1}^3Y_a^2 
- \sum_{A'=1}^{2}\sum_{i=1}^2\left(\Sigma^{A'}H_i - H_i M^{A'}\right) \left(\Sigma^{A'}H_i - H_i M^{A'}\right)^\dagger,
\ee
with
\be
Y_a = e^2\left(c_a - \vec H \sigma_a \vec H^\dagger\right), \qquad
\vec H = \left(H_1,\ H_2\right).
\ee
This is nothing but the ${\cal N}=2$ SQED with $A_\mu$ and $\Sigma^{1,2}$ being bosonic components
of a vector multiplet whereas $H_{i=1,2}$ being those of hypermultiplets (the subscription $i$ is the
index of the $SU(2)_R$ symmetry). 
$Y_{a=1,2}$ are the so-called the Fayet-Illiopoulos (FI) $F$ terms,
and $Y_{a=3}$ is called the FI $D$ term, which form an $SU(2)_R$ triplet.
The constants $c_a$ are the FI-terms
which we can set $c_a = (0,0,v^2)$ without loss of generality.

Now, the SUSY vacua read very similar to those in Eq.~(\ref{eq:vac}) as
\be
\left<A\right> :\quad H_1^B = v \delta^B_A,\qquad H_2^B = 0,\qquad \Sigma^{B'} = m_{B',A}.
\ee
One can easily be convinced that ${\cal L}$ with $D=3$ and $N_F' = 2$ is identical to
${\cal L}^{(D=3)}_{{\cal N}=2}$ except for the additional complex scalar $H_2$.
However, it was found that $H_2$ is completely inert for the BPS states which we are
interested in this paper. Therefore, the Lagrangian (\ref{eq:lag}) with ignoring 
the sterile scalar $H_2$ makes sense (we can include $H_2$ in Eq.~(\ref{eq:lag}) but
it will be identically zero for BPS configurations).


\section{BPS equations for domain wall junctions}
\label{sec:BPS_eq}

From now on, we will investigate BPS states of ${\cal L}$ in Eq.~(\ref{eq:lag})
with $N_F'=D$ under the expectation that the BPS states can exist only when the symmetric structures
in the spatial and the internal spaces are identical as mentioned in Introduction. 
In what follows, the Roman index $m$ stands for the spacial index as $m=1,2,\cdots,D$ and
we will use it for the index of $N_F'$ ($m\equiv A'$).
The energy density of static configurations is
\be
{\cal E} &=& \frac{1}{2e^2}\sum_{m>n}F_{mn}^2
+ \frac{1}{2e^2}\sum_{m,n} (\partial_m\Sigma_n)^2
+ \sum_m D_m H (D_m H)^\dagger \nonumber\\
&\ &+~\frac{1}{2e^2}Y^2
+ \sum_{m}\left(\Sigma_mH - H M_m\right) \left(\Sigma_mH - H M_m\right)^\dagger.
\ee
In order to perform a standard Bogomol'nyi completion to this energy density, let us
first note that the derivative terms of $\Sigma_m$ can be cast into the following form
\be
\sum_{m,n} (\partial_m\Sigma_n)^2 
&=& \sum_m (\partial_m\Sigma_m)^2
+\sum_{m>n}\left\{(\partial_m \Sigma_n)^2 + (\partial_n \Sigma_m)^2\right\} \nonumber\\
&=& \left(\sum_m\xi_m\partial_m\Sigma_m\right)^2
+\sum_{m>n}\left(\partial_m \Sigma_n + \chi_{mn} \partial_n \Sigma_m\right)^2 \nonumber\\
&\ &- 2 \sum_{m>n}\left(\xi_m\xi_n \partial_m\Sigma_m \partial_n\Sigma_n + \chi_{mn}\partial_m \Sigma_n\partial_n \Sigma_m\right),
\ee
where we have introduced the signs $\chi_{mn},\xi_m = \pm 1$.
Then, the Bogomol'nyi completion goes as follows:
\be
{\cal E} &=& \frac{1}{2e^2}\sum_{m>n}\left\{
F_{mn}^2 + \left(\partial_m \Sigma_n + \chi_{mn} \partial_n \Sigma_m\right)^2 \right\}
+ \frac{1}{2e^2}  \left(\sum_m\xi_m\partial_m\Sigma_m - Y\right)^2
\nonumber\\
&\ &+~ \sum_m \left\{D_m H + \xi_m(\Sigma_m H - H M_m)\right\} \left\{D_m H + \xi_m(\Sigma_m H - H M_m)\right\}^\dagger
\nonumber\\
&\ &+~
\sum_m  \xi_m {\cal Z}_m + \sum_{m>n}\xi_m\xi_n {\cal Y}_{mn}+ \sum_m \partial_m {\cal J}_m,
\label{eq:Bog}
\ee
where we have introduced ${\cal Z}_m$, ${\cal Y}_{mn}$, and ${\cal J}_m$ as
\be
{\cal Z}_m &=& v^2\partial_m \Sigma_m, \label{eq:Z_density}\\
{\cal Y}_{mn} &=& - \frac{1}{e^2} \left(\partial_m \Sigma_m \partial_n \Sigma_n + \chi_{mn}\xi_m\xi_n\partial_m\Sigma_n\partial_n\Sigma_m\right),\label{eq:Y_density}\\
{\cal J}_m &=& -\xi_m\left(\Sigma_m H - H M_m\right)H^\dagger.
\ee

The terms ${\cal Z}_m$ and ${\cal Y}_{mn}$ can contribute only topologically. Indeed,
the first quantity ${\cal Z}_m$ is related to a domain wall tension measured along the $x^m$ direction as
\be
Z_m = \int^\infty_{-\infty}dx^m\ \xi_m{\cal Z}_m 
= v^2 \xi_m \bigg(\Sigma_m\big|_{x^m=+\infty} - \Sigma_m\big|_{x^m=-\infty}\bigg).
\label{eq:DW_tension}
\ee
It is well known that $Z_m$ is always {\it positive} regardless of
the choice of $\xi_m$,
which is naturally understood as the domain wall tension.
On the other hand, the contribution of ${\cal J}_m$ vanishes since it is asymptotically zero
because of the vacuum condition (\ref{eq:vac_cond}).
The term ${\cal Y}_{mn}$ is not topological as it is in general. 
To make it topological, we need to impose an additional condition
\be
\chi_{mn}\xi_m\xi_n = -1.
\label{eq:sign}
\ee
Then it becomes
\be
{\cal Y}_{mn}  
= - \frac{1}{e^2} \left(\partial_m \Sigma_m \partial_n \Sigma_n - \partial_m\Sigma_n\partial_n\Sigma_m\right)
= - \frac{1}{e^2}\det\left(
\begin{array}{cc}
\p_m \Sigma_m & \p_m\Sigma_n\\
\p_n \Sigma_n & \p_n \Sigma_n
\end{array}
\right),
\ee
and its integration over the $x^m$-$x^n$ plane gives another topological quantity\footnote{Note
that ${\cal Y}_{mn}$ ($m>n$) can be cast into a total derivative form in the $x^m$-$x^n$ plane as
${\cal Y}_{mn} = -\frac{1}{e^2}\p_k\left(\epsilon^{kl}\Sigma_m\partial_l\Sigma_n\right)$ where $k$ and $l$ take
the value in $m$ and $n$, and $\epsilon^{mn} = -\epsilon^{nm} = 1$.
Therefore, this is a topological charge density.}
\be
Y_{mn} &=&  \xi_m\xi_n \int dx^mdx^n\,{\cal Y}_{mn} = - \frac{\xi_m\xi_n}{e^2} S_{mn},
\ee
where we have defined
\be
S_{mn} \equiv \int dx^mdx^n\, \det\left(
\begin{array}{cc}
\p_m \Sigma_m & \p_m\Sigma_n\\
\p_n \Sigma_n & \p_n \Sigma_n
\end{array}
\right).
\ee
The term $S_{mn}$ corresponds to an area of the region in the $\Sigma^m$-$\Sigma^n$ plane mapped from
the whole $x^m$-$x^n$ plane by the function $(\Sigma_m(x^m,x^n),\Sigma_n(x^m,x^n))$ with all other coordinates
$x^k$ ($k\neq m,n$) being fixed.
Precisely speaking, $S_{mn}$ can be positive or negative, and its absolute value is the area.
Interestingly, $\xi_m\xi_nY_{mn}$ is always {\it negative}\footnote{
The negativeness of $Y_{mn}$ was proved for the $D=2$ case, namely $m=1$ and $n=2$, in Ref.\cite{Eto:2005cp}.
The same proof holds for the generic $Y_{mn}$ in $D \ge 2$, so we do not repeat it here.} 
independent of choice of the signs $\xi_m$ and $\xi_n$,
so it should be understood as a sort of binding energy among domain walls \cite{Oda:1999az,Shifman:1999ri,Ito:2000zf}.

Once we have confirmed that all the terms in the third line of Eq.~(\ref{eq:Bog}) are topological, 
it assures us that the energy density is minimized when the following first order equations are satisfied
\be
F_{mn} = 0,\label{eq:BPS1}\\
\xi_m\partial_m \Sigma_n - \xi_n \partial_n \Sigma_m = 0,\label{eq:BPS2}\\
\sum_m\xi_m\partial_m\Sigma_m - Y = 0,\label{eq:BPS3}\\
\xi_mD_m H + (\Sigma_m H - H M_m) = 0,\label{eq:BPS4}
\ee
where $m,n=1,2,\cdots,D$.
One can verify that all solutions of the above BPS equations solve the full equations of motion.\footnote{Note that if we took the wrong sign $\chi_{mn}\xi_m\xi_n=+1$ 
instead of Eq.~(\ref{eq:sign}), the BPS equations conflict with equations of motion.}
These are a set of the BPS equations of the domain wall junction in $D$ dimensions ($D\ge3$) obtained for the first time. 
Note that they are a $D$ dimensional generalization of the BPS equations of the domain wall junction in $D=2$ cases
studied in Refs.~\cite{Kakimoto:2003zu,Eto:2005cp}.
Note that it is called the $\frac{1}{4}$ BPS equations when $D=2$. It is because a solution of
the equations preserves a quarter of the supersymmetry when embedded 
into a supersymmetric theory Refs.~\cite{Gibbons:1999np,Carroll:1999wr,Gorsky:1999hk}.
Although we have not dealt with any supersymmetry at all in this paper, one can grasp the origin of $\frac{1}{4}$ BPS-ness
by looking at our BPS equations.\footnote{
Indeed, the $\frac{1}{4}$ BPS equations can also be derived via appropriate supersymmetry preserving conditions
to the supersymmetry transformation on the fermionic fields. The sings $\xi_{1,2}$ enter in this argument when 
we selects $\frac{1}{4}$ of the supercharges by two appropriate projection operators.}
When $D=2$, we have two signs $\xi_{1,2} = \pm 1$. 
Therefore, depending on $\xi_{1,2}$,
there exist four different ways for performing the Bogomol'nyi completion to obtain BPS equations.
Turning to our BPS equations in $D$ dimensions, they involve the $D$ signs $\xi_{1,2,\cdots,D}$.
Therefore, there exist $2^D$ different ways to perform the Bogomol'nyi completion of the energy density.


\section{A planar ${\cal S}_3/{\cal S}_2$ domain wall junction in $D=2$: a review}
\label{sec:A3_D2}

\begin{figure}[h]
\begin{center}
\includegraphics[width=13cm]{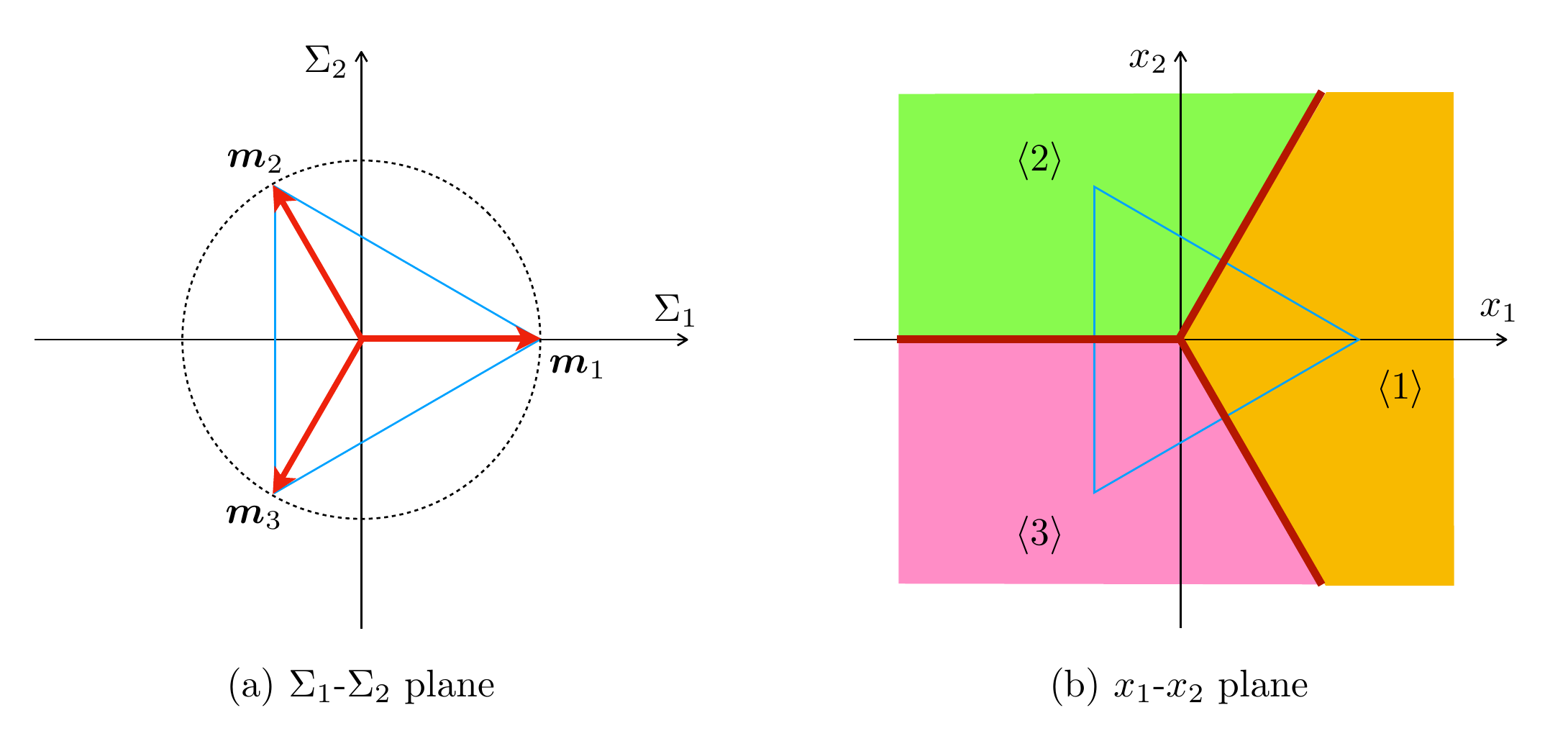}
\caption{(a) The ${\cal S}_3/{\cal S}_2$ mass vectors in the $\Sigma_1$-$\Sigma_2$ plane.
(b) the dual picture depicted in the $x_1$-$x_2$ plane.}
\label{fig:Z3junction}
\end{center}
\end{figure}

As worming up, we briefly review an exact solution for a ${\cal S}_3/{\cal S}_2$ domain wall junction in the case
with $N_F=3$ and $D=N_F'=2$, which was found in Ref.~\cite{Kakimoto:2003zu} (with $N_F=D=3$ and $N_F'=2$).
\footnote{It was called the $\mathbb{Z}_3$ junction in Ref.~\cite{Kakimoto:2003zu}. 
However, note that the symmetry of the Lagrangian is ${\cal S}_3$ but not 
$\mathbb{Z}_3$ and 
that we call it the ${\cal S}_3/{\cal S}_2$ domain wall junction.}  
Now, the indices $m,n$ run from 1 to 2.
The mass matrices are those given in Eqs.~(\ref{eq:mm_Z3}) and (\ref{eq:mv_Z3}), 
leading to the ${\cal S}_3$ symmetric group.
Following the generic argument obtained in Eq.~(\ref{eq:vac}), 
we find three discrete vacua shown in Fig.~\ref{fig:Z3junction}(a). 
In the first vacuum 
$\left<1\right>: \bm{\Sigma} = \bm{m}_1$, the discrete symmetry is spontaneously broken 
as ${\cal S}_3 \to {\cal S}_2$, 
where the elements of the unbroken ${\cal S}_2$ symmetry are given by 
$U_\Sigma \in 
\left\{
\left(\begin{smallmatrix} 1 & 0\\ 0 & 1\end{smallmatrix}\right),\ 
\left(\begin{smallmatrix} 1 & 0\\ 0 & -1\end{smallmatrix}\right)
\right\}$, 
and the corresponding $U_H$ elements are
$U_H \in 
\left\{
\left(\begin{smallmatrix} 1 & 0 & 0\\ 0 & 1 & 0 \\ 0 & 0 & 1\end{smallmatrix}\right),\ 
\left(\begin{smallmatrix} 1 & 0 & 0\\ 0 & 0 & 1\\0 & 1 & 0\end{smallmatrix}\right)
\right\}$.
Thus, the vacuum structure respects the symmetry breaking pattern as ${\cal S}_3/{\cal S}_2$
which corresponds to the three vertices of the equilateral triangle.

To be concrete, we will set $\xi_1 = \xi_2 = +1$ 
without loss of generality 
in what follows.  
The gauge field plays no role in solutions, and so they vanish $A_0 = A_1 = A_2 = 0$.
Then, Eq.~(\ref{eq:BPS1}) is trivially satisfied. 
Eq.~(\ref{eq:BPS2}) is solved by introducing a scalar potential $\phi(x^1,x^2)$ for $\Sigma_{1,2}$ by
\be
\Sigma_{m} = \partial_m \phi,\quad (m=1,2).
\ee
Then, one can verify that Eq.~(\ref{eq:BPS4}) is solved by
\be
H = v e^{-\phi}\left( e^{\bm{m}_1\cdot \bm{x}}
,\ e^{\bm{m}_2\cdot \bm{x}}
,\ e^{\bm{m}_3\cdot \bm{x}}
\right),
\ee
where the mass vectors are those given in Eq.~(\ref{eq:mv_Z3}) and $\bm{x} = (x_1,\ x_2)$.
Note that we have omitted translational moduli parameters since they can always be absorbed by the translational invariance.\footnote{
Although the domain wall junction consists of the three domain walls, the number of the independent moduli are not three but two.
This is because, once we fix the positions of the two domain walls, the position of the third one is automatically determined. Therefore,
we can always fix the moduli by the two-dimensional translations without loss of generality. See Ref.~\cite{Eto:2005cp} for more details.}
Finally, we are left with Eq.~(\ref{eq:BPS3}) which is now expressed in terms of $\phi$ as
\be
\frac{1}{e^2v^2}\triangle_2 \phi = 1-
e^{-2\phi}\psi,
\label{eq:BPS3b}
\ee
where $\triangle_2$ stands for the 2 dimensional Laplacian, and we have introduced a semi-positive function
\be
\psi \equiv
e^{2\bm{m}_1\cdot\bm{x}} + 
e^{2\bm{m}_2\cdot\bm{x}}
+ e^{2\bm{m}_3\cdot\bm{x}}.
\ee
No exact solutions for this equation have been known so far, except for the special case \cite{Kakimoto:2003zu} in which
the model parameters are tuned as
\be
ev = \sqrt{\frac{3}{2}}\,m.
\label{eq:evm_D2}
\ee
In this special case, an exact solution of Eq.~(\ref{eq:BPS3b}) is given by
\be
\phi = \log \left(
e^{\bm{m}_1\cdot\bm{x}} + 
e^{\bm{m}_2\cdot\bm{x}}
+ e^{\bm{m}_3\cdot\bm{x}}
\right).
\label{eq:exact_sol_D2}
\ee
This corresponds to the exact solution found in Ref.~\cite{Kakimoto:2003zu}.

Note that the previous work \cite{Kakimoto:2003zu} for $D=2$ solved the BPS equations in a quite different way 
where the scalar potential was not introduced. 
The procedure of obtaining the solution presented here
is peculiar to the present work. We would like to stress that introducing the scalar potential $\phi$ makes things transparent and 
is an important clue for
obtaining the exact solutions in higher dimensions.

\begin{figure}[h]
\begin{center}
\includegraphics[width=13cm]{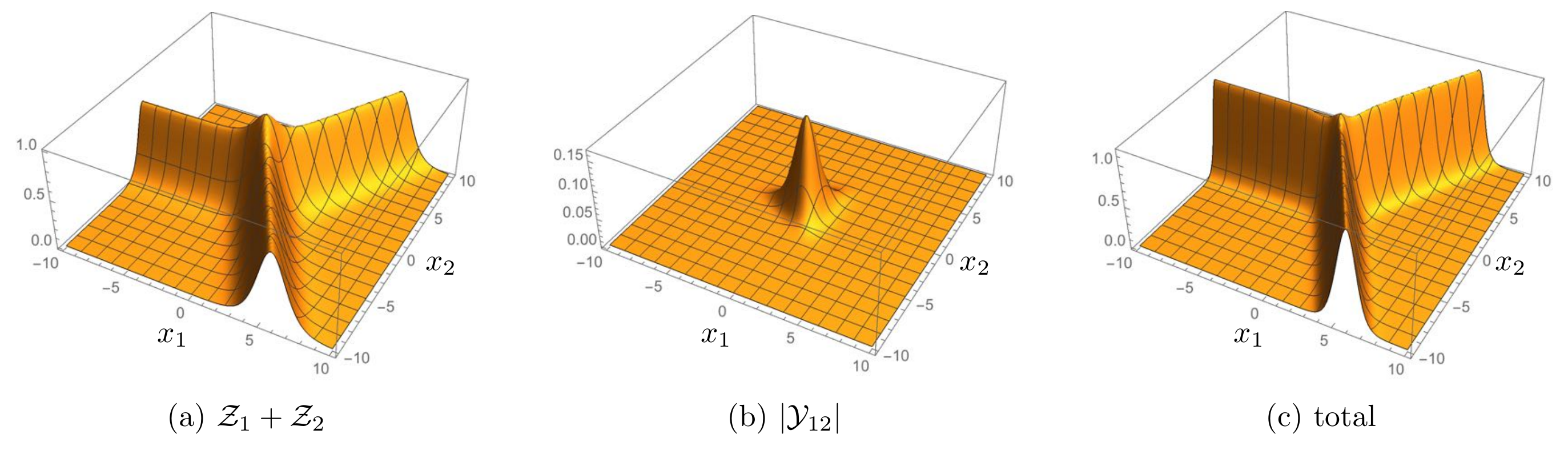}
\caption{The exact solution of the ${\cal S}_3/{\cal S}_2$ planar domain wall junction. 
(a) the domain wall tension density ${\cal Z}_1 + {\cal Z}_2$, (b) 
the absolute value of the ${\cal Y}_{12}$ charge density, and (c) the total energy density.
We set $m=v=1$.
}
\label{fig:A3_D2}
\end{center}
\end{figure}

As a consistency check, let us verify the contribution of ${\cal Y}_{12}$ to the energy density.
It can be expressed as
\be
{\cal E} \supset \xi_1\xi_2 {\cal Y}_{12} = - \frac{1}{e^2} \left[\p_1^2 \phi \p_2^2\phi - (\p_1\p_2\phi)^2\right]
= - \frac{27m^4}{4e^2}e^{-3\phi} < 0.
\ee
Thus, it is negative whole over the $x^1$-$x^2$ plane, and so that its integration is also negative as expected.
Fig.~\ref{fig:A3_D2} shows several plots for the exact solution.

\begin{figure}[h]
\begin{center}
\includegraphics[width=14cm]{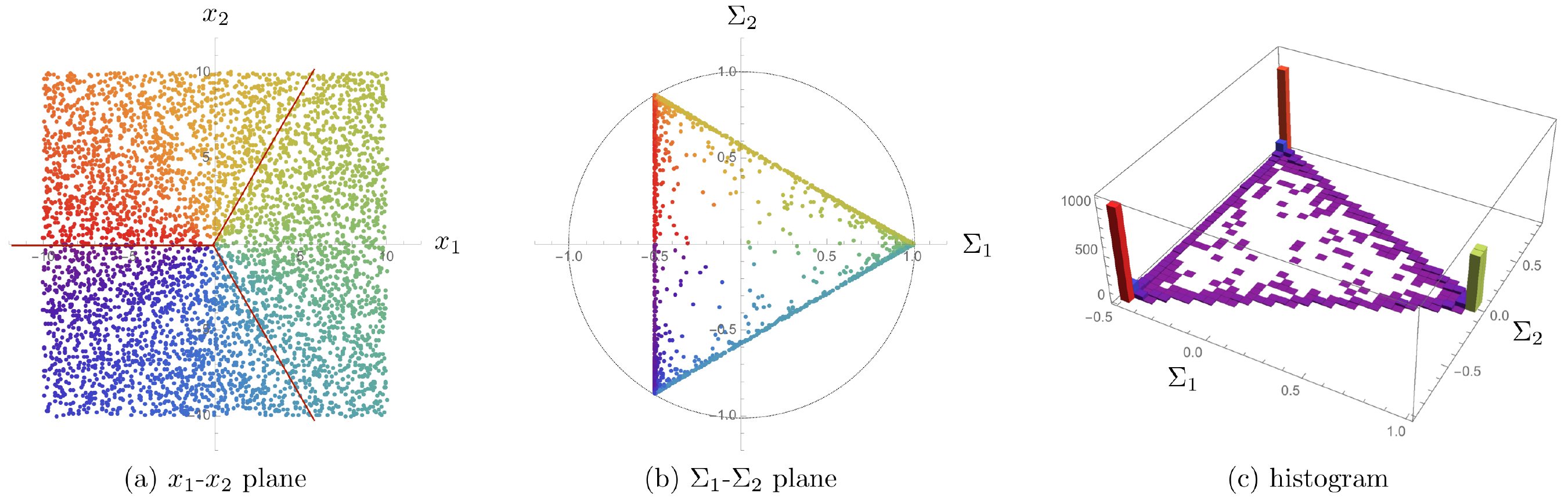}
\caption{
(a) 5000 points are randomly sampled in the region $x_{1,2} \in [-10,10]$, 
(b) the corresponding points in the $\Sigma_1$-$\Sigma_2$ plane are plot, 
and (c) the histogram counting the number of mapped points.
The colors of points correspond to the angle in the $x_1$-$x_2$ plane.
}
\label{fig:A3_D2_mapping}
\end{center}
\end{figure}

It is worthwhile mentioning how useful depicting the solution on the $\Sigma_1$-$\Sigma_2$ plane is.
We should emphasize that 
the Fig.~\ref{fig:Z3junction}(a) has almost all information not only of the vacua but also of $1/2$ BPS
domain walls and $1/4$ BPS domain wall junctions \cite{Eto:2005cp}.
As we explained, the three vertices of the equilateral triangle correspond to the three vacua $\left<A\right>$ with 
$\bm{\Sigma} = \bm{m}_A$ ($A=1,2,3$). In addition, the edge vector $\bm{m}_A - \bm{m}_B$ 
is identical to the domain wall interpolating the vacua $\left<A\right>$ and $\left<B\right>$.
The domain wall tension can easily be read from the length of the edge as
$Z_{AB} = v^2\left|\bm{m}_A - \bm{m}_B\right|$
from Eq.~(\ref{eq:DW_tension}). Finally, the junction of the three domain walls corresponds to
the face of the triangle. As we have explained, the junction charge $Y_{12}$ is proportional to the area
of the triangle, and so it can be easily calculated as $Y_{12} = -\frac{(3+\sqrt3)m^2}{e^2}$. 
These are not all informations we can get from the $\Sigma_1$-$\Sigma_2$ plane, and we can also read
the geometric information of the domain wall junction in the $x^1$-$x^2$ plane. Namely, we can
find angles of the domain walls extending from the junction point: 
they are the orthonormal lines of
the edges of the triangle. 
This can be more rigorously confirmed by regarding $\bm{\Sigma}(x_1,x_2)$ as a function which maps 
the real $x_1$-$x_2$ plane to the internal $\Sigma_1$-$\Sigma_2$ plane.
Fig.~\ref{fig:A3_D2_mapping} shows the image of the mapping: the whole $x_1$-$x_2$ plane is mapped onto
the compact equilateral triangle in the $\Sigma_1$-$\Sigma_2$ plane.
As can be shown in Fig.~\ref{fig:A3_D2_mapping}(c), generic points are mapped onto either of three vertices as expected.
Hence, the configuration in the real space (shown in Fig.~\ref{fig:Z3junction}(b)) is
a dual picture of the internal plane (shown in Fig.~\ref{fig:Z3junction}(b)) \cite{Eto:2005cp}.
They both have the almost same information, and neither of them has a big advantage in the $D=2$ case.
However, as we will see below, the representation in the internal space is easier than the one
in the real space when we go to higher dimensions $D\ge 3$. This is because that the former treats 
compact objects while the latter deals with non-compact configurations.

\section{A tetrahedral ${\cal S}_4/{\cal S}_3$ domain wall junction in $D=3$}
\label{sec:A4_D3}

In this section, we construct a novel exact solution of a non-planar 3D domain wall junction.
To this end, we set $D=N_F' = 3$ ($m=1,2,3)$ and $N_F = 4$ ($A=1,2,3,4$). 
To be concrete, we take $\xi_1 = \xi_2 = \xi_3 = +1$ in what follows.

\begin{figure}[ht]
\begin{center}
\includegraphics[width=13cm]{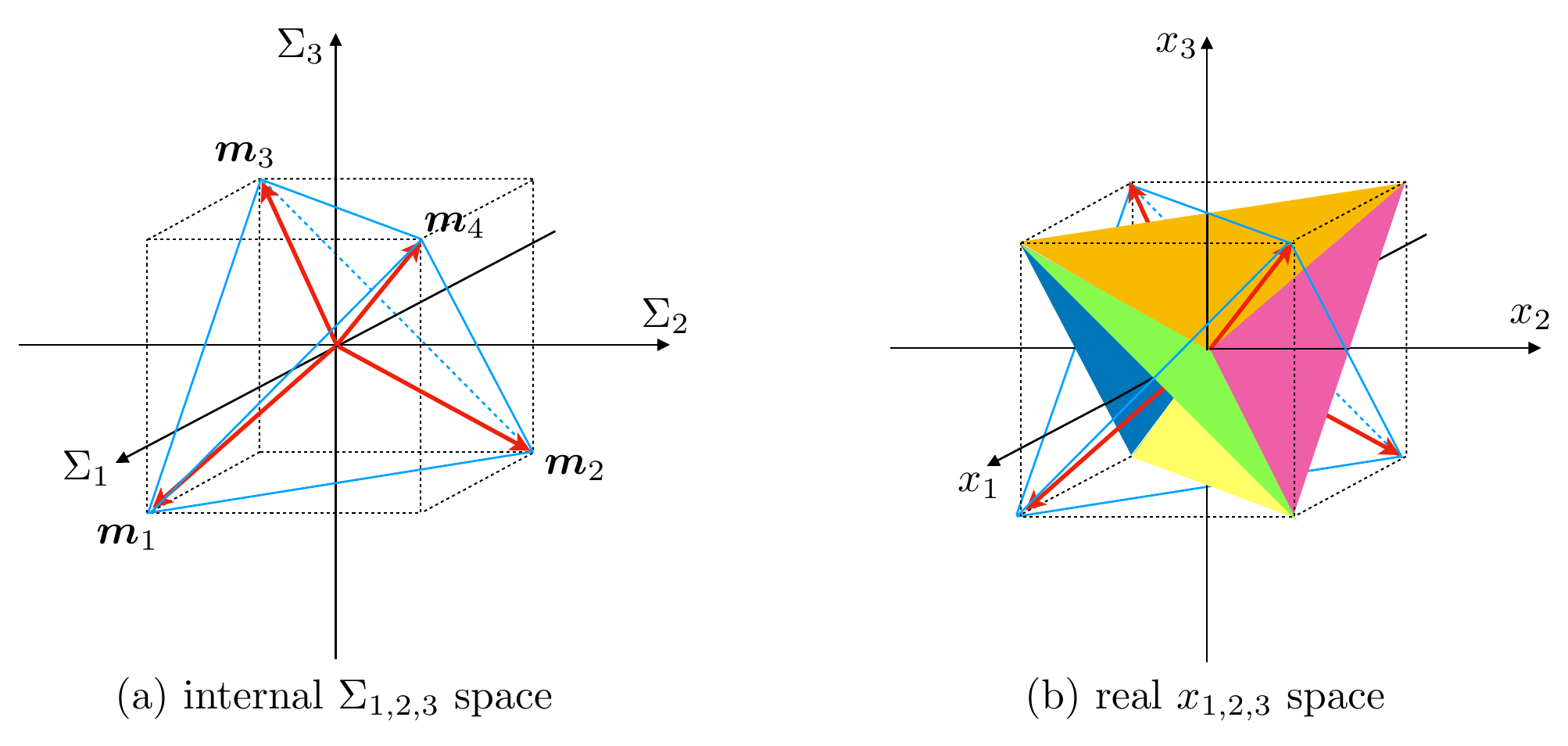}
\caption{(a) The regular tetrahedron in the $\bm{\Sigma}$ space. Each vertex corresponds to the vacuum.
(b) The dual picture of (a) depicted in the real space. The painted triangles correspond to the domain walls.
Only a part of the domain walls inside the auxiliary cube are shown.}
\label{fig:regular_tetrahedron}
\end{center}
\end{figure}

In order to construct an exact solution, we need to arrange the model parameters in such a way
that it has the highest discrete symmetry. Hence, our choice of the mass matrices are of the form
\be
M_1 &=& \frac{m}{\sqrt3}\,{\rm diag}\left(1,\ -1,\ -1,\ 1\right),\label{eq:MM_D3_1}\\
M_2 &=& \frac{m}{\sqrt3}\,{\rm diag}\left(-1,\ 1,\ -1,\ 1\right),\label{eq:MM_D3_2}\\
M_3 &=& \frac{m}{\sqrt3}\,{\rm diag}\left(-1,\ -1,\ 1,\ 1\right),\label{eq:MM_D3_3}
\ee
or equivalently we have
\be
\bm{m}_1 &=& \frac{m}{\sqrt3}\,(1,\ -1,\ -1),\label{eq:m_D3_1}\\ 
\bm{m}_2 &=& \frac{m}{\sqrt3}\,(-1,\ 1,\ -1),\label{eq:m_D3_2}\\ 
\bm{m}_3 &=& \frac{m}{\sqrt3}\,(-1,\ -1,\ 1),\label{eq:m_D3_3}\\ 
\bm{m}_4 &=& \frac{m}{\sqrt3}\,(1,\ 1,\ 1).\label{eq:m_D3_4}
\ee
These are nothing but the four vertices of the regular tetrahedron which are inscribed by a sphere of radius $m$.
Then, the discrete symmetry of the masses is the tetrahedral symmetry which is isomorphic to
the symmetric group of degree four ${\cal S}_4$,
as depicted in Fig.~\ref{fig:regular_tetrahedron}(a).
So, we expect that the
symmetry of the model is also ${\cal S}_4$.
This can be verified by examining the rectangle mass matrix ${\cal M}$
defined in Eq.~(\ref{eq:M}):
\be
{\cal M} = \frac{m}{\sqrt3}\,\left(
\begin{array}{cccc}
1 & -1 & -1 & 1\\
-1 & 1 & -1 & 1\\
-1 & -1 & 1 & 1
\end{array}
\right).
\label{eq:MM_4}
\ee
This is indeed invariant under the transformation 
${\cal M} \to U_\Sigma^T {\cal M} U_H$ with any 4 by 4 matrices $U_H$ of ${\cal S}_4$ together
with the corresponding 3 by 3 matrices $U_\Sigma$ which are explicitly shown in Appendix.
Furthermore, the mass matrices given in Eqs.~(\ref{eq:MM_D3_1}) -- (\ref{eq:MM_D3_3}) satisfy
$M_1^2 + M_2^2 + M_3^2 = m^2{\bf 1}_3$, so that, similarly to Eq.~(\ref{eq:HMMH}),
we have
$
\sum_{A'=1}^3 H M_{A'}{}^2 H^\dagger = m^2 HH^\dagger
$. Hence, all the terms in the Lagrangian
are indeed ${\cal S}_4$ invariant.

Let us mention the vacua. There are four discrete vacua in the model, and selecting one among
four correspond to choosing one vertex from the tetrahedron. 
We choose for instance the first vacuum
$\left<1\right>: \bm{\Sigma} = \bm{m}_1$. 
A subgroup ${\cal S}_3$ of ${\cal S}_4$ which transforms the remaining vertices $\bm{m}_{2,3,4}$ 
is unbroken. Hence, the spontaneous symmetry breaking 
${\cal S}_4 \to {\cal S}_3$ occurs in this case.
The vacuum structure (the discrete four points) respects this and it is isomorphic to
the coset ${\cal S}_4/{\cal S}_3$, see the Appendix for some details.
This is a straightforward extension of ${\cal S}_3/{\cal S}_2$ in the previous subsection.

We are ready for constructing a novel exact solution of a non-planar 3D domain wall junction in the ${\cal S}_4$
symmetric model.
As before, all the gauge fields are set to be zero $A_{\mu=0,1,2,3} = 0$
 so that Eq.~(\ref{eq:BPS1})
is trivially satisfied.
Eq.~(\ref{eq:BPS2}) can be solved by introducing an arbitrary scalar potential $\phi$ as
\be
\Sigma_m = \p_m \phi,\quad (m=1,2,3).
\ee
Then, one can verify that Eq.~(\ref{eq:BPS4}) is solved by
\be
H^A = v\, e^{-\phi} w_A,\qquad
w_A \equiv e^{\bm{m}_A\cdot\bm{x}},\quad (A=1,2,3,4).
\ee
Finally, we are left with Eq.~(\ref{eq:BPS3}) which is now expressed in terms of $\phi$ as
\be
\frac{1}{e^2v^2}\triangle_3 \phi = 1-e^{-2\phi}\psi,
\label{eq:ME_D3}
\ee
where $\triangle_3$ stands for the 3 dimensional Laplacian, and we have introduced a semi-positive function
\be
\psi \equiv
\sum_{A=1}^4 w_A^2.
\ee
Thanks to the ${\cal S}_4$ symmetric masses given in Eqs.~(\ref{eq:m_D3_1}) -- (\ref{eq:m_D3_4}), 
Eq.~(\ref{eq:ME_D3}) can be exactly solved by
\be
\phi = \log \sum_{A=1}^4 w_A,
\ee
only if the model parameters satisfy the following condition
\be
ev = \frac{2}{\sqrt3}\,m.
\label{eq:evm_D3}
\ee
This is the exact analytic solution of a non-planar domain wall junction in $D=3$ obtained for the first time.

Now, we make a comment on the power of the scalar potential $\phi$.
All the procedures to obtain the exact solutions for $D=3$ are straightforward extension of 
those for $D=2$ in the previous section \ref{sec:A3_D2}.
If we tried to solve the BPS equations without the aid of the scalar potential $\phi$ as was done for $D=2$ in Ref.~\cite{Kakimoto:2003zu}, 
it was difficult to reach the BPS equations in $D=3$.\footnote{
Furthermore, the discrete symmetry of the $D=2$ model was
incorrectly understood as $\mathbb{Z}_3$ in Ref.~\cite{Kakimoto:2003zu}. We have found the correct symmetry ${\cal S}_3$ for $D=2$ and 
figured out that the correct extension
of the discrete symmetry is ${\cal S}_4$ for $D=3$, which is also the important clue for discovery of the exact analytic solution in $D=3$.
}
We thus have succeeded in obtaining the new exact solution 
in the case of $D=3$.

Similarly to the planar ${\cal S}_3/{\cal S}_2$ junction in the previous section, we can compute
the ${\cal Y}_{mn}$ charge $(m>n)$ explicitly showing negativeness:
\be
{\cal Y}_{mn} = -\frac{16m^4}{9e^2}
\frac{\displaystyle \sum_{C=1}^4 w_C^{-1}}{\displaystyle \left(\sum_{C=1}^4 w_C\right)^3} < 0,
\ee
for $m,n = 1,2,3$.

\begin{figure}[h]
\begin{center}
\includegraphics[width=15cm]{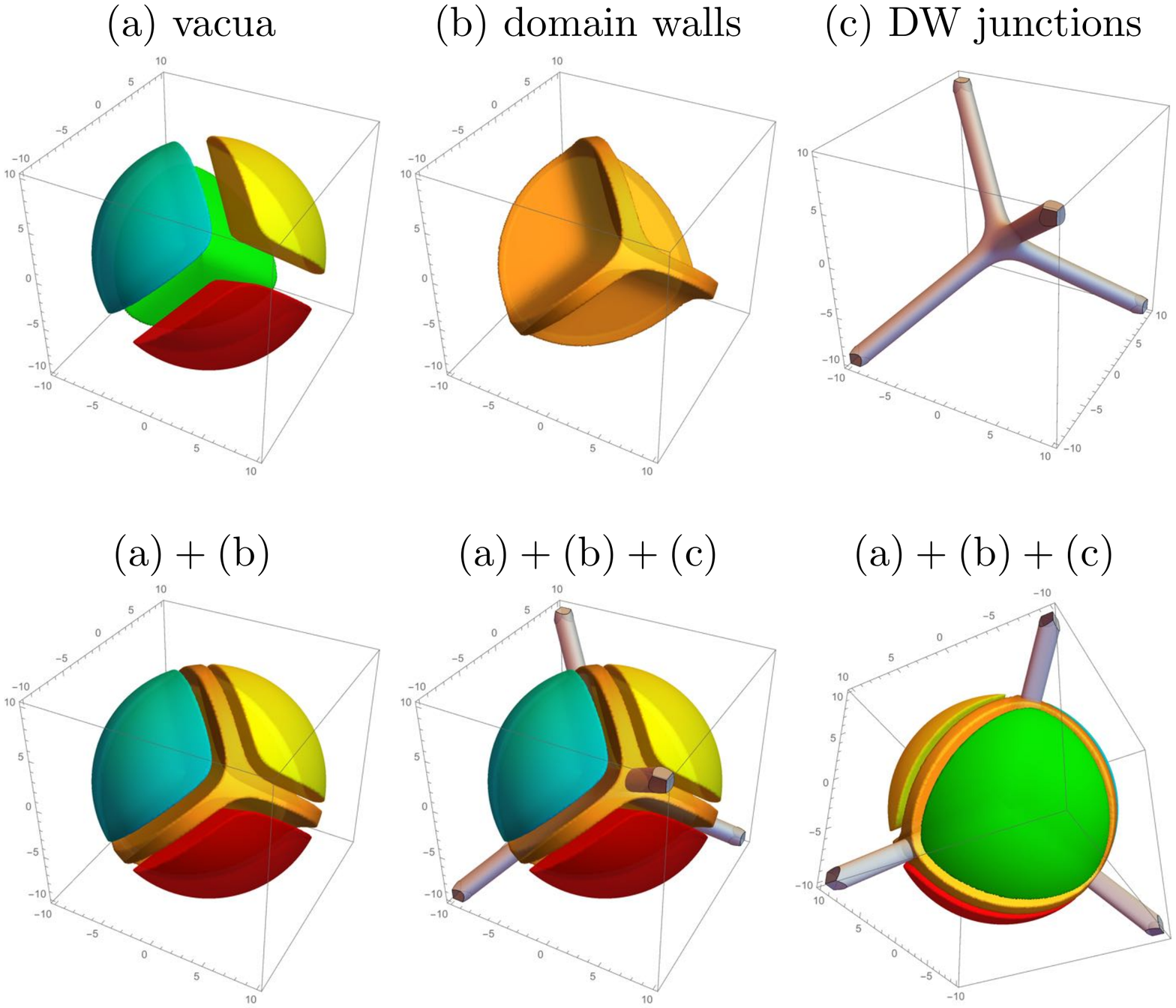}
\caption{The exact solution of the ${\cal S}_4$ tetrahedral domain wall junction. We set $m=v=1$.
(a) shows the four domains of different vacua with $\{\text{red, green, cyan, yellow}\} = \left\{
\left<1\right>,\left<2\right>,\left<3\right>,\left<4\right>\right\}$, 
(b) shows the domain wall energy density $\sum_m{\cal Z}_m$,
(c) shows $\sum_{m>n}{\cal Y}_{mn}$ junction charge density. The panels in the second row
are superpositions of the panels in the first row. The right-most panel shows view from
opposite side of the middle panel.}
\label{fig:A4_D3}
\end{center}
\end{figure}

In Fig.~\ref{fig:A4_D3}, we show several plots of the ${\cal S}_4$ symmetric exact solution in the real space.
For ease of realization of the complicated 3D graphs, we individually plot the vacua, domain walls, and domain wall junctions
in Fig.~\ref{fig:A4_D3}(a)--(c), respectively.
The red region of Fig.~\ref{fig:A4_D3}(a) is identified as the region of
$\{\Sigma_1 > \frac{m}{2} \}\,\land\, \{\Sigma_2 < - \frac{m}{2} \}\,\land\, 
\{\Sigma_3 < -\frac{m}{2}\}$, 
corresponding to the vacuum $\left<1\right>$ with $\bm{\Sigma} = \frac{m}{\sqrt3}(1,-1,-1)$. 
Similarly, the green, cyan, and yellow regions correspond to the vacua $\left<2\right>$,
$\left<3\right>$, and $\left<4\right>$, respectively. 
Note that we only show the interior of the
sphere of the radius $10$ with the unit of $m^{-1}$.
Fig.~\ref{fig:A4_D3}(b) shows a superposition of six domain walls dividing the four vacua, where
the orange part is the region inside which $\sum_{m=1}^3{\cal Z}_m > \frac{1}{2v^2m^2}$ holds.
Fig.~\ref{fig:A4_D3}(c) shows a superposition of four domain wall junctions of the six domain walls,
where the gray part is the region inside which $\sum_{m>n}{\cal Y}_{mn} < - \frac{9}{100} \frac{m^4}{e^2}$ holds.

\begin{figure}[h]
\begin{center}
\includegraphics[width=15cm]{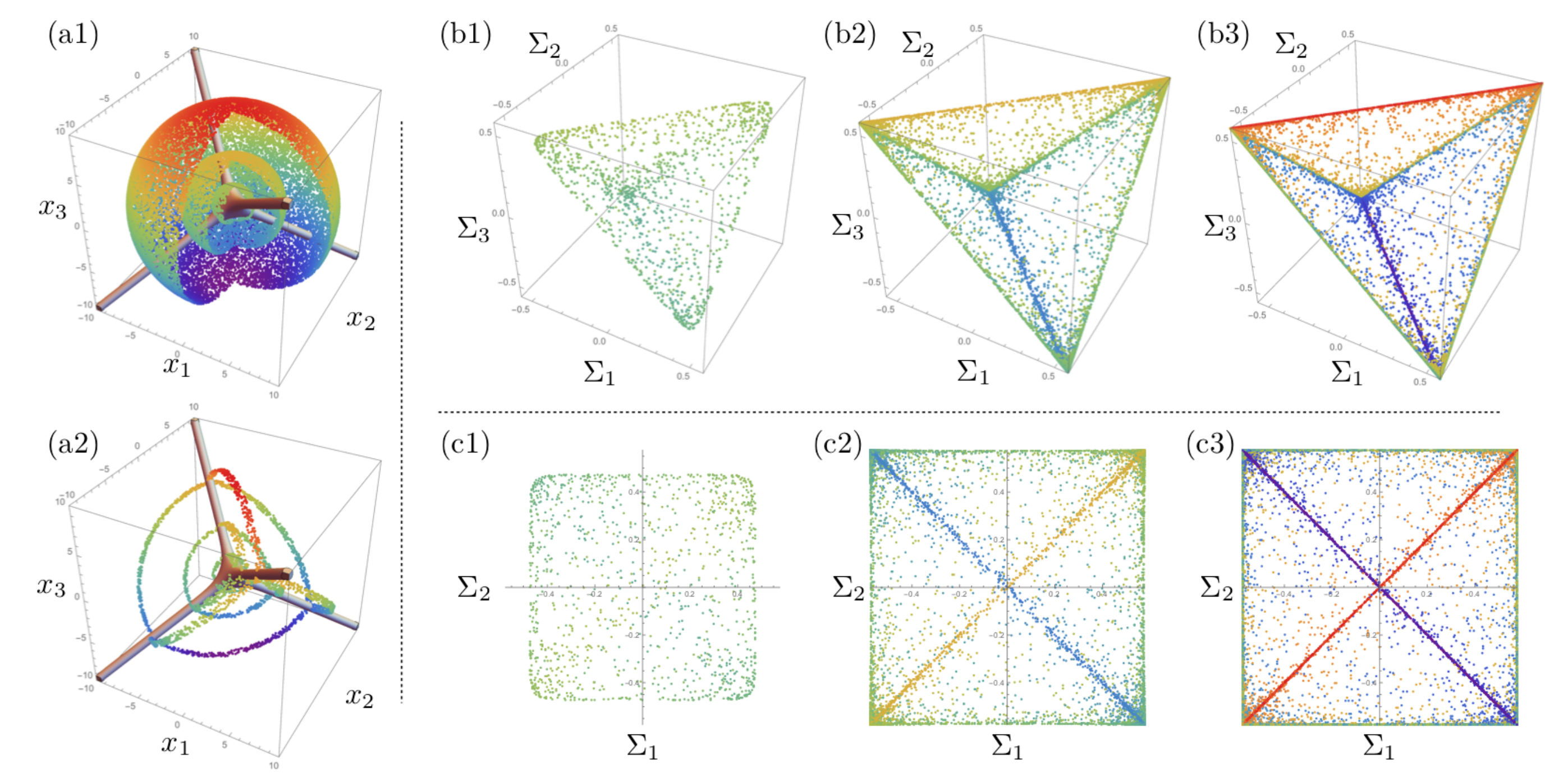}
\caption{(a1) The 1200, 7500, and 30000 sampling points on the spheres of radii $r=2,5,10$.
(a2) The sampling points only inside the domain walls are shown.
(b1)--(b3) The images of the sampling points of (a1).
(c1)--(c3) The orthographic projections of (b1)--(b3) onto the $\Sigma_1$--$\Sigma_2$ plane.}
\label{fig:A4_D3_mapping}
\end{center}
\end{figure}

Let us next describe the exact solution on the internal $\bm{\Sigma}$ plane.
As already shown in Fig.~\ref{fig:regular_tetrahedron}(a), the vacua correspond to
the four vertices of the regular tetrahedron. Then, from our experience with the ${\cal S}_3/{\cal S}_2$ domain wall
junction, we naturally expect that the six edges and four faces correspond to the six domain walls
and the four domain wall junctions, respectively.
To confirm this, we show the map from the internal $\bm{\Sigma}$ space to the real $\bm{x}$ space
by $\bm{\Sigma}(\bm{x})$ in Fig.~\ref{fig:A4_D3_mapping} which is the 3D extension of Fig.~\ref{fig:A3_D2_mapping}.
We take sampling points randomly on the three spheres with radius $r = 2,5,10$ in the real $\bm{x}$ space
as shown in Fig.~\ref{fig:A4_D3_mapping}(a1). The number of the sampling points are 1200, 
7500, and 30000, respectively. Their images on the internal $\bm{\Sigma}$ space are shown in 
Fig.~\ref{fig:A4_D3_mapping}(b1)--(b3). We can see that generic points on the spheres are mapped onto either of
the vertices of the tetrahedron. The population on the edges are much less than those on the vertices
since inverse images of points on the edges are arcs [Fig.~\ref{fig:A4_D3_mapping}(a2)] 
which are intersections of the sampling spheres and the domain walls.
The sampling points inside the domain wall junctions are mapped onto the faces of the tetrahedron.
Clearly, there are much rare points on the faces. Comparing (b1), (b2) and (b3), one sees that 
the tetrahedral shape becomes clearer for the lager sphere. 
We also show orthographic projections of the tetrahedral images 
onto the $\Sigma_1$--$\Sigma_2$ plane as Fig.~\ref{fig:A4_D3_mapping}(c1)--(c3).

\begin{figure}[h]
\begin{center}
\includegraphics[width=12cm]{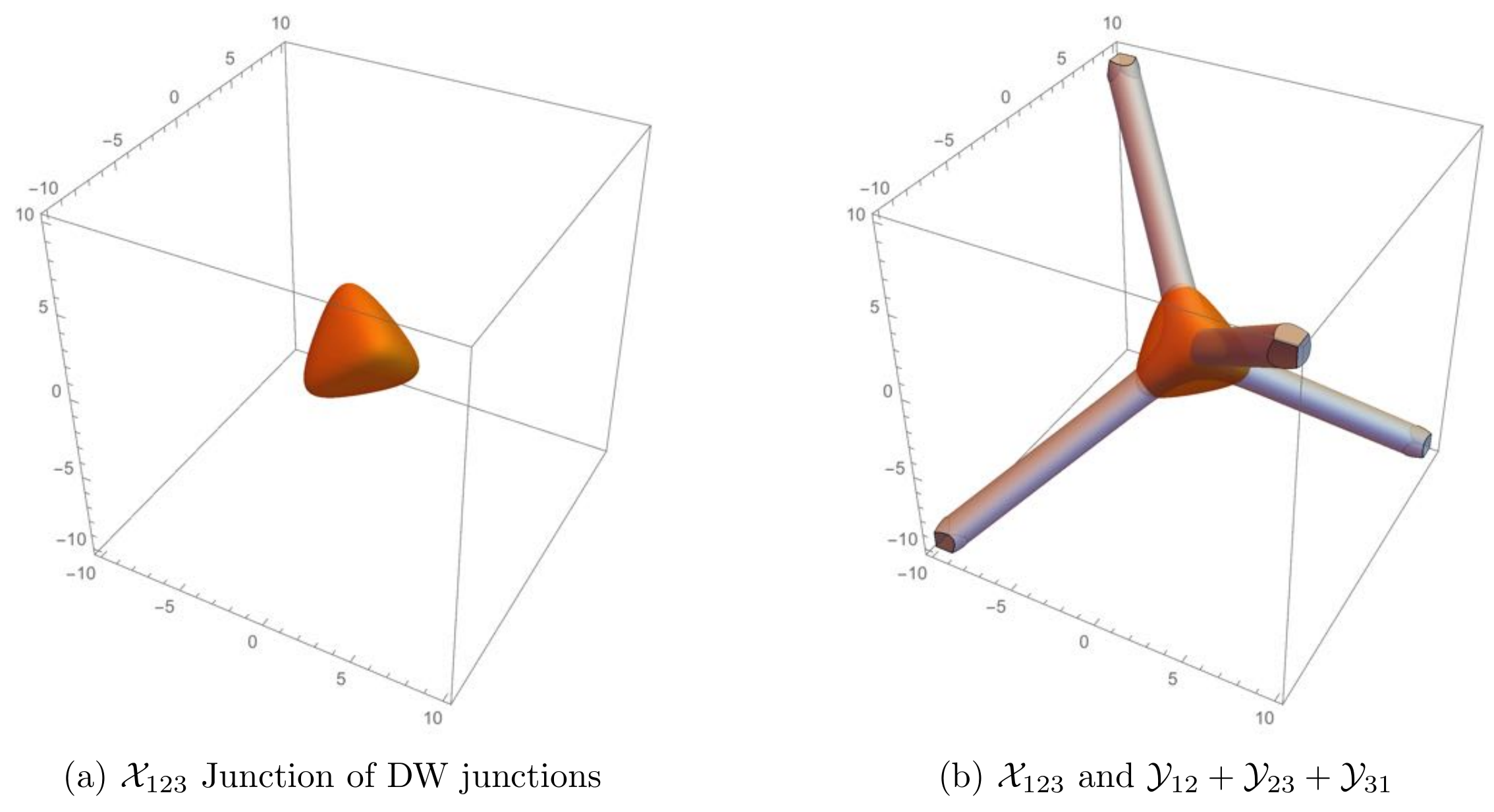}
\caption{The topological charge density ${\cal X}_{123}$ of the junction of the domain wall junctions.
(a) shows the constant-level surface with ${\cal X}_{123}$, and (b) shows superposition of Fig.~\ref{fig:A4_D3_DWJJ}(a) and Fig.~\ref{fig:A4_D3}(c).}
\label{fig:A4_D3_DWJJ}
\end{center}
\end{figure}

As we have learned above, there is a good relation between the configuration expressed
in the $\bm{x}$ space and that in the $\bm{\Sigma}$ space. As is summarized in Tab.~\ref{tab:1},
the vacua, domain walls, and the domain wall junctions correspond to the 3D domains,
the 2D planes, and the 1D lines in the $\bm{x}$ space, and to the vertices,
the edges, and the faces of the tetrahedron in the $\bm{\Sigma}$ space.
At this point, we realize that there is one more piece, namely the junction 
of the domain wall junctions in the $\bm{x}$ space, which is a point-like object.
Its natural counterpart in the $\bm{\Sigma}$ space should be 3D interior of the tetrahedron.
\begin{table}[ht]
\begin{center}
\begin{tabular}{|c||c|c|c|c|}
\hline
 & vac & DW & DW junction & junction of DW junctions\\
\hline
$\bm{x}$ space & 3D domain & plane (2D) & line (1D) & point (0D)\\
$\bm{\Sigma}$ space & vertex (0D) & edge (1D) & face (2D) & interior (3D)\\
degree & 4 & 6 & 4 & 1\\
\hline
\end{tabular}
\caption{Duality among the constituents of the ${\cal S}_4/{\cal S}_3$ three-dimensional domain wall junction 
in the $\bm{x}$ space and the $\bm{\Sigma}$ space.}
\label{tab:1}
\end{center}
\end{table}
This can be already seen in Fig.~\ref{fig:A4_D3_mapping}(b1) where the small sphere near
the junction of the domain wall junctions is mapped on the tetrahedral inner surface.
In order to make a more rigorous argument, we return to the Bogomol'nyi completion in Eq.~(\ref{eq:Bog}). 
There, we have found that the two topological quantities ${\cal Z}_m$ and ${\cal Y}_{mn}$ 
for the domain walls and the domain wall junctions, respectively.
No other topological objects take part. However, we now realize that ${\cal Z}_m$ is the 1D
Jacobian between $x_m$ and $\Sigma_m$, and ${\cal Y}_{mn}$ is the 2D Jacobian between
$\{x_m,x_n\}$ and $\{\Sigma_m,\Sigma_n\}$. Along this line, we naturally define a new topological
quantity as the 3D dimensional Jacobian between $\{x_1,x_2,x_3\}$ 
and $\{\Sigma_1,\Sigma_2,\Sigma_3\}$ by
\be
{\cal X}_{123} \equiv \det \left(
\begin{array}{ccc}
\p_1 \Sigma_1 & \p_1 \Sigma_2 & \p_1 \Sigma_3\\
\p_2 \Sigma_1 & \p_2 \Sigma_2 & \p_2 \Sigma_3\\
\p_3 \Sigma_1 & \p_3 \Sigma_2 & \p_3 \Sigma_3
\end{array}
\right).
\label{eq:X}
\ee
Note that this can be cast into the total derivative form as
\be
{\cal X}_{123} = \p_l\left(\epsilon^{lmn}\Sigma_1\p_m\Sigma_2\p_n\Sigma_3\right),
\ee
with $m,n,l$ runs from 1 to 3. Therefore, this is a topological charge density.
For the ${\cal S}_4$ symmetric solution, we have
\be
X = \int d^3x\ {\cal X}_{123} = \frac{1}{3}\left(\frac{2m}{\sqrt3}\right)^3.
\ee
This corresponds to the volume of the tetrahedron in the $\bm{\Sigma}$ space.
Fig.~\ref{fig:A4_D3_DWJJ} shows a 3D constant-level surface with ${\cal X}_{123} = \frac{1}{100}$,
which is indeed a tetrahedral blowup of the junction point 
of the domain wall junctions in the $\bm{x}$ space.


\section{Exact solutions of ${\cal S}_{D+1}/{\cal S}_D$ domain wall junctions}
\label{sec:simpler}

\subsection{Deriving exact solutions in generic dimensions}
\label{sec:exact_sol_D}

The derivation of the exact solutions of the domain wall junctions 
in Secs.~\ref{sec:A3_D2} and \ref{sec:A4_D3} can be generalized straightforwardly to higher dimensions.
To this end, let us consider the $(D+1)$-dimensional model with $N_F' = D$ and $N_F = D+1$
(we have studied $D=2,3$ in 
Secs.~\ref{sec:A3_D2} and \ref{sec:A4_D3}).

We will prove 
that the $\frac{1}{2^D}$ BPS equations (\ref{eq:BPS1}) -- (\ref{eq:BPS4})
admit an exact solution
\be
A_\mu = 0,\qquad 
\Sigma_m = \partial_m \phi,\qquad H_A = v e^{-\phi} w_A,
\label{eq:sol_D_1}
\ee
where $m=1,2,\cdots,D$ and $A=1,2,\cdots,D+1$, and $w_A$ and $\phi$ are given by
\be
w_A = e^{\bm{m}_A\cdot\bm{x}},\qquad \phi = \log \sum_{A=1}^{D+1}w_A,
\label{eq:sol_D_2}
\ee
and the mass vectors $\bm{m}_A$ should 
correspond to the coordinates of $D+1$ vertices of a regular $D$-simplex of the radius
$m$, namely $|\bm{m}_A| = m$. 
One can easily show that Eqs.~(\ref{eq:BPS1}), (\ref{eq:BPS2}), and (\ref{eq:BPS4})
are solved by Eqs.~(\ref{eq:sol_D_1}) and (\ref{eq:sol_D_2}). Then,
it will turn out that a highlight of the proof is verifying Eq.~(\ref{eq:BPS3})
which is written as
\be
\frac{1}{e^2v^2}\triangle_D^2 \phi = 1-e^{-2\phi}\psi,\qquad
\psi \equiv
\sum_{A=1}^{D+1} w_A^2,
\label{eq:ME_D}
\ee
only if the relation
\be
ev =  \sqrt{\frac{D+1}{D}} m
\label{eq:para_D}
\ee 
holds. This is the first solution for the domain wall junction in generic $D$ dimensions.

\begin{figure}[h]
\begin{center}
\includegraphics[width=14cm]{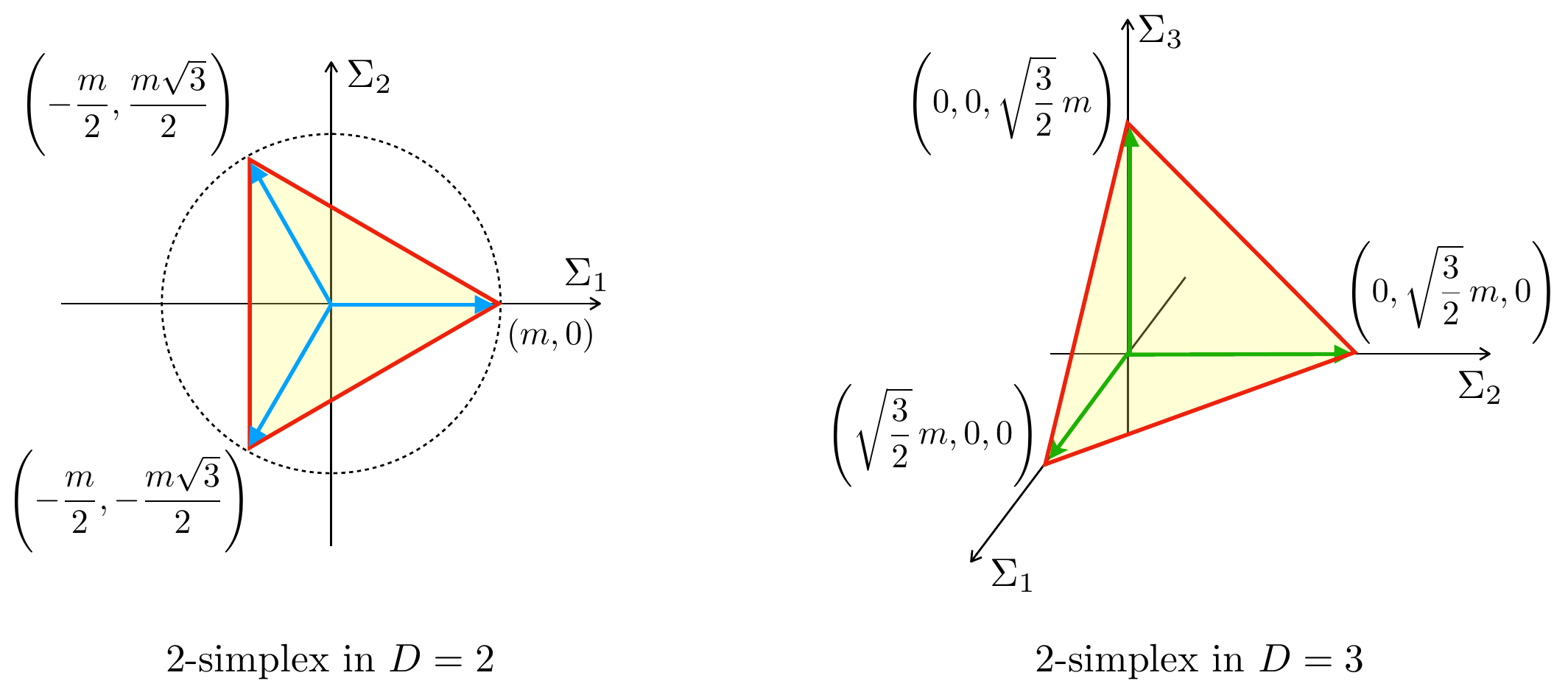}
\caption{A regular 2-simplex (an equilateral triangle) drawn in $D=2$ and $D=3$ spaces.}
\label{fig:2-simplex}
\end{center}
\end{figure}

One of the less beautiful parts in the following argument comes from complexity 
of giving the coordinate of the
each vertex of the regular $D$-simplex in the internal $D$ dimensional space,
see for example Eq.~(\ref{eq:mv_Z3}) for $D=2$ 
and Eqs.~(\ref{eq:m_D3_1})--(\ref{eq:m_D3_4}) for $D=3$.
In order to avoid this inessential complexity, 
we here embed the problem of the $D$ dimensional real and internal spaces into
the $D+1$ dimensional ones. 
Namely, we consider ${\cal L}$ with the spacetime dimensions $(D+1)+1$ and the flavor number $N_F'=D+1$,  
with keeping $N_F=D+1$. Namely,
we leave the $D$-simplex as it is (we do not consider a $D+1$-simplex),
see Fig.~\ref{fig:2-simplex} for the case of 2-simplex before and after the embedding.
There are two advantages to do this. 
One is that
the mass vectors (corresponding to the vertices of the regular $D$-simplex) can be simply expressed as
\be
\bm{m}_1 &=& m' \left(1,0,0,\cdots,0,0\right),\label{eq:D+1_simplex_1}\\
\bm{m}_2 &=& m' \left(0,1,0,\cdots,0,0\right),\\
&\vdots&\nonumber\\
\bm{m}_D &=& m' \left(0,0,0,\cdots,1,0\right),\\
\bm{m}_{D+1} &=& m' \left(0,0,0,\cdots,0,1\right),\label{eq:D+1_simplex_D+1}
\ee
with 
\be
m' = \sqrt{\frac{D+1}{D}}\,m.
\ee
The second merit is that 
the discrete symmetry among these vectors is clearly the ${\cal S}_{D+1}$ group
which is identical to permutations of the $D+1$ axes.
Thanks to the first advantage, $w_A$ becomes drastically simple as
\be
w_A = e^{m'x^A},\quad (A=1,2,\cdots,D+1).
\ee
Then, we can easily verify that $\phi$ given in Eq.~(\ref{eq:sol_D_2}) satisfies the following equation
\be
\triangle_{D+1} \phi 
= m'{}^2\left(1 - e^{-2\phi}\psi\right).
\label{eq:ME_D+1}
\ee
This is identical to Eq.~(\ref{eq:ME_D}) with Eq.~(\ref{eq:para_D}) 
only except for the Laplacian $\triangle_{D+1}$
instead of $\triangle_D$.
To complete the proof, we need to show the equivalence between
Eqs.~(\ref{eq:ME_D}) and (\ref{eq:ME_D+1}). This can be done as follows.
Firstly, we note that any $D$-simplex is included in a $D$ dimensional subspace of the $D+1$
dimensional space.
However, we used the $D+1$ vectors $\bm{m}_A$ ($A=1,2,\cdots,D+1)$ 
for expressing the $D+1$ vertices of the regular $D$-simplex
as in Eqs.~(\ref{eq:D+1_simplex_1}) -- (\ref{eq:D+1_simplex_D+1}). Therefore, there exists an appropriate
$U \in SO(D+1)$ transformation which transforms the vectors as
\be
\bm{m}_A \to \tilde{\bm{m}}_A = U\bm{m}_A=m'\left( \tilde m_{1,A},\ \cdots,\ \tilde m_{D,A},\ 0\right),\quad \text{for all }A.
\ee
At the same time, we can also redefine the extended $D+1$ dimensional spacial coordinate by
the same $SO(D+1)$ element by
\be
\bm{x} \to \tilde{\bm{x}} = U \bm{x}.
\ee
This transformation does not change the inner product
\be
\bm{m}_A\cdot\bm{x} = \tilde{\bm{m}}_A^T U^T U \tilde{\bm{x}} = \tilde{\bm{m}}_A\cdot\tilde{\bm{x}}.
\label{eq:coordnate_change}
\ee
Since the $(D+1)$th component of $\tilde{\bm{m}}_A$ $(A=1,2,\cdots,D+1)$ is zero, the $(D+1)$th coordinate $\tilde x_{D+1}$
does not appear in Eq.~(\ref{eq:coordnate_change}).
Both $\phi$ and $\psi$ include $\tilde{\bm x}$ only through the combination 
$\bm{m}_A\cdot\bm{x}= \tilde{\bm{m}}_A\cdot\tilde{\bm{x}}$, so that
$\tilde x_{D+1}$ is completely redundant.
Hence, the $(D+1)$ dimensional Laplacian in terms of the new coordinate $\tilde{\bm x}$ 
is identical to the $D$ dimensional one in
Eq.~(\ref{eq:ME_D+1}).

We again have faced the power of the scalar potential $\phi$.
All the procedures to obtain the exact analytic solutions for generic $D$ are straightforward extension of 
those developed in this work for $D=2$ in the previous section \ref{sec:A3_D2}, together with
the correct understanding of the discrete symmetry ${\cal S}_{D+1}$. Without these developments, we could not find the analytic solutions.


\subsection{Geometric properties}
\label{sec:geo_pro}

We now have the picture that the ${\cal S}_{D+1}/{\cal S}_D$ domain wall junction
in $D$ dimensional real space is dual to the regular $D$-simplex.
Due to this correspondence, we can easily understand the physical structure of the higher dimensional domain wall junctions. 
In mathematics, the convex hull of a subset of size $d+1$ of the $D+1$ points of the $D$-simplex is called
a $d$-face of the $D$-simplex (the 0-faces are the vertices, the 1-faces are the edges,
and the $(D-1)$-faces are called the facets).
The number of the $d$-faces is the binomial coefficient 
$\left(\begin{smallmatrix}D+1\\d+1\end{smallmatrix}\right)$.
Several low-lying examples are shown in Tab.~\ref{tab:2}.
In the physics context, by the duality, we relate the 0-faces to the $D$-dimensional vacuum domains, 
the 1-faces to the $(D-1)$-dimensional domain walls, and the 2-faces to the $(D-2)$-dimensional
domain wall junctions, and so on. We call the $d$-faces the $d$-walls.
Namely, the 0-wall means the vacuum, the 1-wall corresponds to the domain wall, and
the 2-wall stands for the domain wall junction, and so on.
We have already studied the 2- and 3-simplexes in Sec.~\ref{sec:A3_D2} and \ref{sec:A4_D3},
respectively.
\begin{table}[ht]
\begin{center}
\begin{tabular}{|c||c|c|c|c|c|c|c|}
\hline
 & 0-face & 1-face & 2-face & 3-face & 4-face & 5-face & 6-face\\
\hline
1-simplex & 2 & 1 &  & & & & \\
2-simplex & 3 & 3 & 1 & & & & \\
3-simplex & 4 & 6 & 4 & 1 & & & \\
4-simplex & 5 & 10 & 10 & 5 & 1& & \\
5-simplex & 6 & 15 & 20 & 15 & 6 & 1 & \\
6-simplex & 7 & 21 & 35 & 35 & 21 & 7 & 1\\
\hline
 & 0-wall & 1-wall & 2-wall & 3-wall & 4-wall & 5-wall & 6-wall\\
\hline
\end{tabular}
\caption{The number of  $d$-faces ($d$-walls) of  $D$-simplex for $D=1,2,3,4,5,6$.}
\label{tab:2}
\end{center}
\end{table}

Let us introduce a topological quantity to each of the $d$-wall for $d\ge1$.
For $d=1$, the natural quantity is ${\cal Z}_m$ defined in Eq.~(\ref{eq:Z_density})
and its integration over the $x_m$ coordinate. 
Similarly, we ran into ${\cal Y}_{mn}$ in Eq.~(\ref{eq:Y_density}) and its integral
over the $x_m$-$x_n$ plane for the 2-wall  for $d=2$.
Note that the numbers of ${\cal Z}_m$ and ${\cal Y}_{mn}$ of the ${\cal S}_{D+1}/{\cal S}_D$ domain
wall junction are $D$ and 
$\frac{(D-1)D}{2!}$, respectively, which coincide with
$\left(\begin{smallmatrix}D\\d\end{smallmatrix}\right)$ for $d=1$ and $d=2$, respectively.
For $d=3$, we have encountered  ${\cal X}_{123}$ given in Eq.~(\ref{eq:X})
for the ${\cal S}_4/{\cal S}_3$ domain wall junction in the model with $D=3$. 
Having ${\cal Z}_m$ and ${\cal Y}_{mn}$ for generic ${\cal S}_{D+1}/{\cal S}_D$ domain wall junction
together with ${\cal X}_{123}$ of the ${\cal S}_4/{\cal S}_3$ domain wall junction at hand, we are now ready to
define the topological quantity of the level $d$ by
\be
{\cal W}_d(m_1,m_2,\cdots,m_d) = \det
\left(
\begin{array}{cccc}
\p_{m_1}\Sigma_{m_1} & \p_{m_1}\Sigma_{m_2} & \cdots & \p_{m_1}\Sigma_{m_d}\\
\p_{m_2}\Sigma_{m_1} & \p_{m_2}\Sigma_{m_2} & \cdots & \p_{m_2}\Sigma_{m_d}\\
\vdots & & \ddots & \vdots \\
\p_{m_d}\Sigma_{m_1} & \p_{m_d}\Sigma_{m_2} & \cdots & \p_{m_d}\Sigma_{m_d}
\end{array}
\right),
\label{eq:W_d}
\ee
where $m_\alpha \in \{1,2,\cdots,D\}$ ($\alpha=1,2,\cdots,d$) 
and $m_\alpha > m_\beta$ if $\alpha > \beta$.
The number of the level-$d$ ${\cal W}$ is $\left(\begin{smallmatrix}D\\d\end{smallmatrix}\right)$,
see Tab.~\ref{tab:3} for several low-lying $D=1,2,3,4,5,6$.
Note that this can be cast into the total derivative form. For example, ${\cal W}_D$ can be expressed as
\be
{\cal W}_D = \p_{m_1}\left(\epsilon^{m_1\cdots m_D}\Sigma_1\p_{m_2}\Sigma_2\p_{m_3}\Sigma_3 \cdots \p_{m_D}\Sigma_D\right).
\ee
Thus, this is a topological charge density. The same can be said for ${\cal W}_d$ for all $d \le D$.
\begin{table}[ht]
\begin{center}
\begin{tabular}{|c||c|c|c|c|c|c|}
\hline
 & level-1 ${\cal W}$ & level-2 ${\cal W}$ & level-3 ${\cal W}$ & level-4 ${\cal W}$ & level-5 ${\cal W}$ & level-6 ${\cal W}$ \\
\hline
${\cal S}_2$ DW & 1 &  &  & & &  \\
${\cal S}_3/{\cal S}_2$ DWJ & 2 & 1 &  & & &  \\
${\cal S}_4/{\cal S}_3$ DWJ & 3 & 3 & 1 & & &  \\
${\cal S}_5/{\cal S}_4$ DWJ & 4 & 6 & 4 & 1 & &  \\
${\cal S}_6/{\cal S}_5$ DWJ & 5 & 10 & 10 & 5 & 1 &  \\
${\cal S}_7/{\cal S}_6$ DWJ & 6 & 15 & 20 & 15 & 6 & 1 \\
\hline
 & ${\cal Z}_m$ & ${\cal Y}_{mn}$ & ${\cal X}_{lmn}$ &  &  & \\
\hline
\end{tabular}
\caption{The number of level-$d$ ${\cal W}$ of ${\cal S}_{D+1}/{\cal S}_D$ domain wall junction for $D=1,2,3,4,5,6$.}
\label{tab:3}
\end{center}
\end{table}

By definition, the level-$d$ wall charge coincides with the volume of a $d$-face of the $D$-simplex
inscribed by a $D$-sphere of radius $m$ 
[the side length is $\sqrt{2(D+1)/D}\,m$
defined in Eqs.~(\ref{eq:D+1_simplex_1})--(\ref{eq:D+1_simplex_D+1})]
\be
W_d(m_1,m_2,\cdots,m_d) &=& \int dx_{m_1}dx_{m_2}\cdots dx_{m_d}\, \left|{\cal W}(m_1,m_2,\cdots,m_d)\right|\nonumber\\
&=& \left(\sqrt{\frac{2(D+1)}{D}}m\right)^d \frac{\sqrt{d+1}}{d!2^\frac{d}{2}}.
\ee
The level-1 ${\cal W}_1(x_m)$ and the level-2 ${\cal W}_2(x_m,x_n)$ 
are related to ${\cal Z}_m$ and ${\cal Y}_{mn}$, respectively.

\begin{figure}[h]
\begin{center}
\includegraphics[width=14cm]{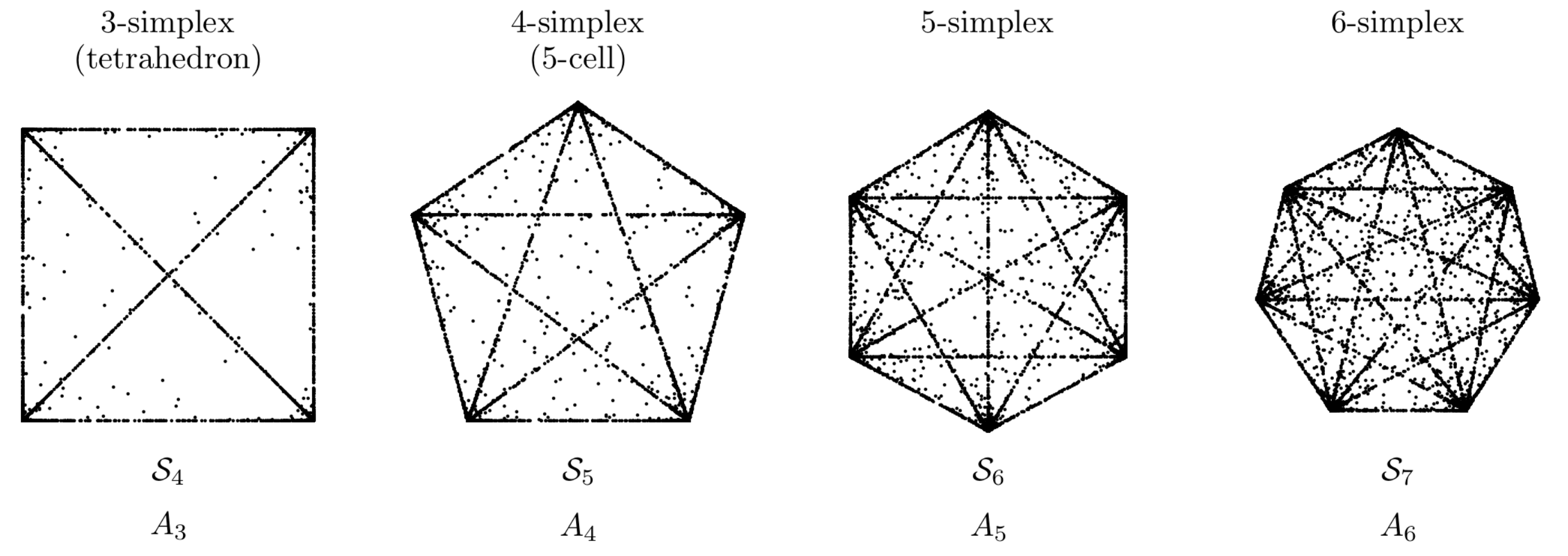}
\caption{Orthographic projections of images for randomly chosen 20,000 points on a $(D-1)$-sphere
(radius $|\bm{x}|=40$) by the exact solutions
of ${\cal S}_{D+1}/{\cal S}_D$ domain wall junctions with $D=3,4,5,6$. The diagrams are identical to the
so called the Coxeter plane of type $A_D$ for a regular $D$-simplex.}
\label{fig:coxeter_plane}
\end{center}
\end{figure}

Before closing this section, let us briefly sketch a higher dimensional junctions which
we have not shown yet. The first example is the ${\cal S}_5/{\cal S}_4$ domain wall junction (the 4-simplex)
in the model with $N_F=5$ and $D=N_F'=4$. 
A concrete set of the five mass vectors are given by
\be
\bm{m}_1 &=& \frac{\sqrt{5}}{4}m\left(1,-1,-1,-\frac{1}{\sqrt5}\right),\\
\bm{m}_2 &=& \frac{\sqrt{5}}{4}m\left(-1,1,-1,-\frac{1}{\sqrt5}\right),\\
\bm{m}_3 &=& \frac{\sqrt{5}}{4}m\left(-1,-1,1,-\frac{1}{\sqrt5}\right),\\
\bm{m}_4 &=& \frac{\sqrt{5}}{4}m\left(1,1,1,-\frac{1}{\sqrt5}\right),\\
\bm{m}_5 &=& \frac{\sqrt{5}}{4}m\left(0,0,0,\frac{4}{\sqrt5}\right).
\ee
The corresponding ${\cal S}_5/{\cal S}_4$ domain wall junction divides
$D=4$ space into the 5 vacuum domains. 
It is not easy to imagine such non-compact higher dimensional object, but, instead, 
all the geometric data can be easy read from Tab.~\ref{tab:2}.
As a supplementary, we show the images on a two dimensional plane in $\bm{\Sigma}$ space
whose preimages are
randomly chosen 20,000 points on the 3-sphere $S^3$ of the radius $|\bm{x}| = 40$ in
Fig.~\ref{fig:coxeter_plane}. 
Almost all points are mapped on either of five vertices (0-walls)
of the pentagon,
and the much less points are mapped on the ten edges (1-walls) of the pentagon and pentagram.  
Much rarer points are mapped onto interior of the pentagon. These two dimensional images, the pentagon
and pentagram, are familiar for the 5-cell. It is a orthographic projection of the 5-cell onto
a 2 dimensional plane (so called the Coxeter plane). The pentagon with pentagram is the type $A_4$
Coxeter plane in which all the vertices and edges of 5-cells are separately shown. The symbol $A_4$
comes from the fact that the symmetric group ${\cal S}_5$ of the 5-cell is isomorphic to the
Coxeter group of the type $A_4$.

The symmetry group ${\cal S}_{D+1}$ of a regular $D$-simplex is known 
as the Coxeter group of type $A_D$, and its Coxeter plane is known as 
a convex regular $(D+1)$-gon. 
Our exact solution of the ${\cal S}_{D+1}/{\cal S}_D$ domain wall junction gives
the Coxeter plane of type $A_D$ via the map from the $D$-dimensional real space $\bm{x}$ to
the $D$-dimensional internal space $\bm{\Sigma}$. We show several concrete diagrams for $D=3,4,5,6$
in Fig.~\ref{fig:coxeter_plane}.


\section{Summary and discussion}
\label{sec:conclusion}

In this work, we have constructed the exact solutions of the ${\cal S}_{D+1}/{\cal S}_D$ domain wall junctions
in $D+1$-dimensional spacetime.
We have considered SUSY motivated Abelian gauge theories 
with $N_F$ charged complex scalar fields $H_A$ and $N_F'=D$ real scalar fields $\Sigma_{A'}$, and have derived the new
BPS equations for the domain wall junctions.
We then have restricted ourselves to the cases with specific flavor numbers $N_F=D+1$, 
and have obtained the analytic exact BPS solutions in the extended Abelian-Higgs model for the first time.
There are two necessary conditions for finding the exact solutions.
The first is the ${\cal S}_{D+1}$ symmetric masses $\bm{m}_A$ 
so that the mass vectors
should be placed at vertices of a regular $D$-simplex in the internal space. 
The other is the special relation
between the coupling constants given in Eq.~(\ref{eq:para_D}).
When these conditions are satisfied, we have found that the exact BPS solutions can be 
obtained by the scalar potential $\phi$ as in Eqs.~(\ref{eq:sol_D_1}) and (\ref{eq:sol_D_2}).
We should emphasize that introducing the scalar potential $\phi$ has been crucial to reformulate the
complicated BPS equations to be surprisingly simple. We have been able to accomplish constructing the analytic solutions
with the aid of $\phi$.
We have verified that the solution for $D=2$ is identical to the one previously obtained in Ref.~\cite{Kakimoto:2003zu} in which
the scalar potential $\phi$ was not used.
We also have developed how to describe such non-compact extended solitons in higher dimensional spaces.
We have found that there is a one-to-one correspondence between 
the BPS configuration in the real $\bm{x}$ space and that in the internal $\bm{\Sigma}$ space.
The latter is more useful and tractable because it deals with compact objects, the regular $D$-simplex.
This correspondence allows us to understand what kind of intersections are included.
We have found that all the building blocks are in one-to-one
correspondence to the $d$-faces (we called them the $d$-walls) 
of the regular $D$-simplex.
Thus, the ${\cal S}_{D+1}/{\cal S}_D$ domain wall junction consists of 
$\left(\begin{smallmatrix}D+1\\d+1\end{smallmatrix}\right)$ $d$-walls, see Tab.~\ref{tab:2}.
The Bogomol'nyi completion shows that the energy density of the ${\cal S}_{D+1}$
domain wall junctions depends on only two kinds of the topological charge densities, 
the domain wall (1-wall) charge $Z_m$ and the domain wall junction (2-wall)
charge $Y_{mn}$.
The topological charge densities 
${\cal Z}_m$ and ${\cal Y}_{mn}$ are a 1-dimensional Jacobian 
from $x_m$ to $\Sigma_m$ 
and  a 2-dimensional Jacobian from $x_m$-$x_n$ to $\Sigma_m$-$\Sigma_n$, 
respectively. 
Then, we have been naturally lead to extend them for the generic $d$-walls by
the $d$-dimensional Jacobian from the real $\mathbb{R}^d$ space to the internal 
$\mathbb{R}^d$ space. The simplest example is $X$ ($d=3$) for the tetrahedral 
domain wall junction ($D=3$), see Fig.~\ref{fig:A4_D3_DWJJ}.
All the topological charges for the generic $d$-walls have been unified by
the level-$d$ wall charge $W_d$ in Eq.~(\ref{eq:W_d}). 
The final achievement in this work is on the visualization of the ${\cal S}_{D+1}/{\cal S}_D$ domain wall junctions. 
Describing the regular $D$-simplex in lower dimensions 
is known as the Coxeter plane. It is a two-dimensional projection on
a convex regular $(D+1)$-gon which is 
 the Coxeter diagram of type $A_D$. 
We have found that the exact solutions $\bm{\Sigma}(\bm{x})$
are canonical mappings from the former to the regular $D$-simplex
in the $\bm{\Sigma}$ space, and their images are nothing but the Coxeter diagram of the type $A_D$.

We have assumed the special relation among the parameters $e, v$, and the masses, in order to
find the analytic solutions of the newly found BPS equations in generic $D$ dimensions.
The models are limited, however,
with the analytic solutions at our hands, now we clearly have understood the whole picture of the domain wall junctions
in higher dimensions. We have been able to easily explain how many $d$-walls are included in the $D$ dimensional junction. 
Since these properties are topological, they do not change even when we continuously deform the model parameters.
For generic parameters, there are no analytic solutions and therefore one needs to solve numerically the complicated differential
equations in $D$ dimensions.

Before closing this paper, 
let us mention several future directions.
First, we have considered only ${\cal S}_{D+1}$ symmetric masses corresponding to vertices of the regular $D$-simplex to construct the exact solutions.
If we consider more general masses corresponding to a deformed $D$-simplex, the corresponding the domain wall junction is also deformed
from the ${\cal S}_{D+1}$ symmetric configuration. 
For such generic case, we cannot expect 
the existence of analytic solutions for generic couplings, 
but in the strong coupling limit analytic solutions would be available  
as for the $D=2$ case \cite{Eto:2005cp}.
For generic dimensions, we have exhausted all exact solutions with full moduli parameters in a separated paper \cite{Eto:2020cys}.
More generally,  we would need numerical works to obtain the BPS solutions 
for generic couplings. 

Second, we have studied the minimal models in which the flavor number of the charged scalar fields is chosen as $N_F = D+1$. 
When we increase the number $N_F$ of flavors, 
the number of the vacua increases accordingly. 
This leads to generic polytopes other than the $D$-simplex which correspond to
more complicated networks in the real $\bm{x}$ space. 
In particular, it admits domain wall loops for $D=2$ \cite{Eto:2005cp}, 
in which case the low energy effective action 
was constructed for localized modes of domain wall loops 
\cite{Eto:2006bb} 
and the low energy dynamics of such loops was 
studied in the moduli approximation \cite{Eto:2007uc}.
The same can be done for 
higher dimensional extensions in this paper, 
which would admits higher dimensional loops, like holes surrounded by domain walls.

Third,
we can extend the Abelian gauge theory studied in this paper to $U(N_C)$ gauge 
theories. 
For the $D=2$ case, it was found that the non-Abelian extension of the junction charge
${\cal Y}_{12}$ exist in the non-Abelian gauge theories \cite{Eto:2005cp}.
In such a case, if some masses are degenerated, 
there appear non-Abelian moduli on the domain walls, 
called non-Abelian clouds
\cite{Shifman:2003uh,Eto:2005cc,Eto:2008dm}. 
Such non-Abelian moduli would also appear in the domain wall junctions 
in general dimensions.

Fourth, we can consider non-BPS domain wall junction in $D+1$-dimensional spacetime, 
while  in this paper,
we have concentrated only on the BPS solutions. 
Once we relax the BPS condition, we would have
more generic networks of the domain walls, as for the simplest 
Wess-Zumino model \cite{Saffin:1999au}.

Fifth, as for a possible connection with differential geometry, 
the case of $D=2$ was found to be interpreted 
in terms of toropical geometry and amoeba in mathematics
\cite{Fujimori:2008ee}.
Higher dimensional extensions of the present paper 
will provide higher dimensional correspondence 
to toropical geometry and amoeba.

Finally, let us mention possible applications of our model to physics.
Single or parallel domain walls can be applied to Josephson junction arrays 
of superconductors sandwiching  insulators 
\cite{Nitta:2012xq,Fujimori:2016tmw}. 
This can be extended to 3-dimensional junction 
found in this paper.
Also, cosmological domain wall networks 
\cite{cosmology} 
will be one of the most interesting physical applications.


\section*{Acknowledgements}

This work was supported by the 
Ministry of Education, Culture, 
Sports, Science (MEXT)-Supported Program for the Strategic Research 
Foundation at Private Universities ``Topological  Science'' 
(Grant  No.~S1511006) 
and JSPS KAKENHI Grant Numbers 16H03984.
The work of M.~E.~ is also supported in part 
by JSPS Grant-in-Aid for Scientific Research 
KAKENHI Grant No. JP19K03839, and
by MEXT KAKENHI Grant-in-Aid for 
Scientific Research on Innovative Areas
``Discrete Geometric Analysis for Materials Design'' No. JP17H06462 from the MEXT of Japan.
The work of M.~N.~is also supported 
in part by JSPS KAKENHI Grant Number 18H01217 
by 
a Grant-in-Aid for Scientific Research on Innovative Areas 
``Topological Materials Science'' 
(KAKENHI Grant No.~15H05855) from MEXT of Japan.

\begin{appendix}

\section{The symmetric group ${\cal S}_4$ and the coset ${\cal S}_4/{\cal S}_3$}
\label{sec:app}
We give the 4 by 4 matrix representation of ${\cal S}_4$:
\be
U_H \in 
\left\{
\begin{array}{c}
\left(
\begin{smallmatrix}
1 & 0 & 0 & 0\\
0 & 1 & 0 & 0\\
0 & 0 & 1 & 0\\
0 & 0 & 0 & 1
\end{smallmatrix}
\right),\ 
\left(
\begin{smallmatrix}
1 & 0 & 0 & 0\\
0 & 1 & 0 & 0\\
0 & 0 & 0 & 1\\
0 & 0 & 1 & 0
\end{smallmatrix}
\right),\ 
\left(
\begin{smallmatrix}
1 & 0 & 0 & 0\\
0 & 0 & 1 & 0\\
0 & 1 & 0 & 0\\
0 & 0 & 0 & 1
\end{smallmatrix}
\right),\ 
\left(
\begin{smallmatrix}
1 & 0 & 0 & 0\\
0 & 0 & 0 & 1\\
0 & 1 & 0 & 0\\
0 & 0 & 1 & 0
\end{smallmatrix}
\right),\
\left(
\begin{smallmatrix}
1 & 0 & 0 & 0\\
0 & 0 & 1 & 0\\
0 & 0 & 0 & 1\\
0 & 1 & 0 & 0
\end{smallmatrix}
\right),\  
\left(
\begin{smallmatrix}
1 & 0 & 0 & 0\\
0 & 0 & 0 & 1\\
0 & 0 & 1 & 0\\
0 & 1 & 0 & 0
\end{smallmatrix}
\right),\\
\left(
\begin{smallmatrix}
0 & 1 & 0 & 0\\
1 & 0 & 0 & 0\\
0 & 0 & 1 & 0\\
0 & 0 & 0 & 1
\end{smallmatrix}
\right),\ 
\left(
\begin{smallmatrix}
0 & 1 & 0 & 0\\
1 & 0 & 0 & 0\\
0 & 0 & 0 & 1\\
0 & 0 & 1 & 0
\end{smallmatrix}
\right),\ 
\left(
\begin{smallmatrix}
0 & 0 & 1 & 0\\
1 & 0 & 0 & 0\\
0 & 1 & 0 & 0\\
0 & 0 & 0 & 1
\end{smallmatrix}
\right),\ 
\left(
\begin{smallmatrix}
0 & 0 & 0 & 1\\
1 & 0 & 0 & 0\\
0 & 1 & 0 & 0\\
0 & 0 & 1 & 0
\end{smallmatrix}
\right),\ 
\left(
\begin{smallmatrix}
0 & 0 & 1 & 0\\
1 & 0 & 0 & 0\\
0 & 0 & 0 & 1\\
0 & 1 & 0 & 0
\end{smallmatrix}
\right),\ 
\left(
\begin{smallmatrix}
0 & 0 & 0 & 1\\
1 & 0 & 0 & 0\\
0 & 0 & 1 & 0\\
0 & 1 & 0 & 0
\end{smallmatrix}
\right),\\
\left(
\begin{smallmatrix}
0 & 1 & 0 & 0\\
0 & 0 & 1 & 0\\
1 & 0 & 0 & 0\\
0 & 0 & 0 & 1
\end{smallmatrix}
\right),\ 
\left(
\begin{smallmatrix}
0 & 1 & 0 & 0\\
0 & 0 & 0 & 1\\
1 & 0 & 0 & 0\\
0 & 0 & 1 & 0
\end{smallmatrix}
\right),\ 
\left(
\begin{smallmatrix}
0 & 0 & 1 & 0\\
0 & 1 & 0 & 0\\
1 & 0 & 0 & 0\\
0 & 0 & 0 & 1
\end{smallmatrix}
\right),\ 
\left(
\begin{smallmatrix}
0 & 0 & 0 & 1\\
0 & 1 & 0 & 0\\
1 & 0 & 0 & 0\\
0 & 0 & 1 & 0
\end{smallmatrix}
\right),\ 
\left(
\begin{smallmatrix}
0 & 0 & 1 & 0\\
0 & 0 & 0 & 1\\
1 & 0 & 0 & 0\\
0 & 1 & 0 & 0
\end{smallmatrix}
\right),\ 
\left(
\begin{smallmatrix}
0 & 0 & 0 & 1\\
0 & 0 & 1 & 0\\
1 & 0 & 0 & 0\\
0 & 1 & 0 & 0
\end{smallmatrix}
\right),\\
\left(
\begin{smallmatrix}
0 & 1 & 0 & 0\\
0 & 0 & 1 & 0\\
0 & 0 & 0 & 1\\
1 & 0 & 0 & 0
\end{smallmatrix}
\right),\ 
\left(
\begin{smallmatrix}
0 & 1 & 0 & 0\\
0 & 0 & 0 & 1\\
0 & 0 & 1 & 0\\
1 & 0 & 0 & 0
\end{smallmatrix}
\right),\ 
\left(
\begin{smallmatrix}
0 & 0 & 1 & 0\\
0 & 1 & 0 & 0\\
0 & 0 & 0 & 1\\
1 & 0 & 0 & 0
\end{smallmatrix}
\right),\ 
\left(
\begin{smallmatrix}
0 & 0 & 0 & 1\\
0 & 1 & 0 & 0\\
0 & 0 & 1 & 0\\
1 & 0 & 0 & 0
\end{smallmatrix}
\right),\ 
\left(
\begin{smallmatrix}
0 & 0 & 1 & 0\\
0 & 0 & 0 & 1\\
0 & 1 & 0 & 0\\
1 & 0 & 0 & 0
\end{smallmatrix}
\right),\ 
\left(
\begin{smallmatrix}
0 & 0 & 0 & 1\\
0 & 0 & 1 & 0\\
0 & 1 & 0 & 0\\
1 & 0 & 0 & 0
\end{smallmatrix}
\right)
\end{array}
\right\},
\label{eq:A4_rep4}
\ee 
and the corresponding 3 by 3 representation:
\be
U_\Sigma \in
\left\{
\begin{array}{c}
\left(
\begin{smallmatrix}
1 & 0 & 0 \\
0 & 1 & 0 \\
0 & 0 & 1 
\end{smallmatrix}
\right),\ 
\left(
\begin{smallmatrix}
0 & -1 & 0 \\
-1 & 0 & 0 \\
0 & 0 & 1 
\end{smallmatrix}
\right),\ 
\left(
\begin{smallmatrix}
1 & 0 & 0 \\
0 & 0 & 1 \\
0 & 1 & 0 
\end{smallmatrix}
\right),\ 
\left(
\begin{smallmatrix}
0 & -1 & 0 \\
0 & 0 & 1 \\
-1 & 0 & 0 
\end{smallmatrix}
\right),\ 
\left(
\begin{smallmatrix}
0 & 0 & -1 \\
-1 & 0 & 0 \\
0 & 1 & 0 
\end{smallmatrix}
\right),\ 
\left(
\begin{smallmatrix}
0 & 0 & -1 \\
0 & 1 & 0 \\
-1 & 0 & 0 
\end{smallmatrix}
\right),\\
\left(
\begin{smallmatrix}
0 & 1 & 0 \\
1 & 0 & 0 \\
0 & 0 & 1 
\end{smallmatrix}
\right),\ 
\left(
\begin{smallmatrix}
-1 & 0 & 0 \\
0 & -1 & 0 \\
0 & 0 & 1 
\end{smallmatrix}
\right),\ 
\left(
\begin{smallmatrix}
0 & 0 & 1 \\
1 & 0 & 0 \\
0 & 1 & 0 
\end{smallmatrix}
\right),\ 
\left(
\begin{smallmatrix}
0 & 0 & 1 \\
0 & -1 & 0 \\
-1 & 0 & 0
\end{smallmatrix}
\right),\ 
\left(
\begin{smallmatrix}
-1 & 0 & 0 \\
0 & 0 & -1 \\
0 & 1 & 0 
\end{smallmatrix}
\right),\ 
\left(
\begin{smallmatrix}
0 & 1 & 0 \\
0 & 0 & -1 \\
-1 & 0 & 0 
\end{smallmatrix}
\right),\\
\left(
\begin{smallmatrix}
0 & 1 & 0 \\
0 & 0 & 1 \\
1 & 0 & 0 
\end{smallmatrix}
\right),\ 
\left(
\begin{smallmatrix}
-1 & 0 & 0 \\
0 & 0 & 1 \\
0 & -1 & 0 
\end{smallmatrix}
\right),\ 
\left(
\begin{smallmatrix}
0 & 0 & 1 \\
0 & 1 & 0 \\
1 & 0 & 0 
\end{smallmatrix}
\right),\ 
\left(
\begin{smallmatrix}
0 & 0 & 1 \\
-1 & 0 & 0 \\
0 & -1 & 0 
\end{smallmatrix}
\right),\ 
\left(
\begin{smallmatrix}
-1 & 0 & 0 \\
0 & 1 & 0 \\
0 & 0 & -1 
\end{smallmatrix}
\right),\ 
\left(
\begin{smallmatrix}
0 & 1 & 0 \\
-1 & 0 & 0 \\
0 & 0 & -1 
\end{smallmatrix}
\right),\\
\left(
\begin{smallmatrix}
0 & 0 & -1 \\
0 & -1 & 0 \\
1 & 0 & 0 
\end{smallmatrix}
\right),\ 
\left(
\begin{smallmatrix}
0 & 0 & -1 \\
1 & 0 & 0 \\
0 & -1 & 0 
\end{smallmatrix}
\right),\ 
\left(
\begin{smallmatrix}
0 & -1 & 0 \\
0 & 0 & -1 \\
1 & 0 & 0 
\end{smallmatrix}
\right),\ 
\left(
\begin{smallmatrix}
1 & 0 & 0 \\
0 & 0 & -1 \\
0 & -1 & 0 
\end{smallmatrix}
\right),\ 
\left(
\begin{smallmatrix}
0 & -1 & 0 \\
1 & 0 & 0 \\
0 & 0 & -1 
\end{smallmatrix}
\right),\ 
\left(
\begin{smallmatrix}
1 & 0 & 0 \\
0 & -1 & 0 \\
0 & 0 & -1 
\end{smallmatrix}
\right)
\end{array}
\right\}.
\label{eq:A4_rep3}
\ee
These satisfy the relation (\ref{eq:discrete_sym}) with the mass matrix ${\cal M}$ given
in Eq.~(\ref{eq:MM_4}). Note that the one to one correspondence between $U_H$ and $U_\Sigma$
is given by
$U_\Sigma = \left({\cal M}{\cal M}^T\right)^{-1}{\cal M} U_H^{-1T} {\cal M}^T$.
The $4!=24$ matrices of Eq.~(\ref{eq:A4_rep4}) coincide 
to the standard 4 by 4 representation of ${\cal S}_4$.
On the contrary, the 3 by 3 matrices in Eq.~(\ref{eq:A4_rep3}) might not be familiar, but indeed
they also form another representation of ${\cal S}_4$. 

The six matrices in the top row of Eq.~(\ref{eq:A4_rep4}) which is a subgroup ${\cal S}_3$ in ${\cal S}_4$
leave the first vacuum intact as
\be
\bm{\Sigma}\big|_{\left<1\right>} = \bm{m}_1 = \frac{m}{\sqrt3}\left(\begin{array}{c}1\\-1\\-1\end{array}\right)
\quad \to \quad U_\Sigma \bm{m}_1 = \bm{m}_1.
\ee
Thus, the coset of ${\cal S}_4$ by the ${\cal S}_3$ quotient has four representatives
\be
{\cal S}_4/{\cal S}_3 = \left\{
\left(
\begin{smallmatrix}
1 & 0 & 0 \\
0 & 1 & 0 \\
0 & 0 & 1 
\end{smallmatrix}
\right),\ 
\left(
\begin{smallmatrix}
0 & 1 & 0 \\
1 & 0 & 0 \\
0 & 0 & 1 
\end{smallmatrix}
\right),\ 
\left(
\begin{smallmatrix}
0 & 1 & 0 \\
0 & 0 & 1 \\
1 & 0 & 0 
\end{smallmatrix}
\right),\ 
\left(
\begin{smallmatrix}
0 & 0 & -1 \\
0 & -1 & 0 \\
1 & 0 & 0 
\end{smallmatrix}
\right)\right\}.
\ee
Note that ${\cal S}_3$ is not a normal subgroup of ${\cal S}_4$. Therefore,
the coset ${\cal S}_4/{\cal S}_3$ is not a group.

\end{appendix}

\bibliographystyle{jhep}

\end{document}